\documentclass[a4paper,fleqn,usenatbib]{mnras}
\usepackage{graphicx}
\usepackage{amssymb}
\usepackage{amsmath}
\usepackage{ae,aecompl}
\usepackage{natbib,times}
\usepackage{lscape}

\newcommand{\mc}{\multicolumn}

\begin{document}
\title[Galaxy clusters in DES$\times$unWISE] {Clusters of galaxies up to
  $z=1.5$ identified from photometric data of the Dark Energy
  Survey and unWISE}

\author[Wen \& Han]
{Z. L. Wen$^{1,2}$\thanks{E-mail: zhonglue@nao.cas.cn}
  and J. L. Han$^{1,2,3}$
\\
1. National Astronomical Observatories, Chinese Academy of Sciences, 
20A Datun Road, Chaoyang District, Beijing 100012, China\\
2. CAS Key Laboratory of FAST, NAOC, Chinese Academy of Sciences,
           Beijing 100101, China \\
3. School of Astronomy, University of Chinese Academy of Sciences,
           Beijing 100049, China 
}

\date{Accepted XXX. Received YYY; in original form ZZZ}

\label{firstpage}
\pagerange{\pageref{firstpage}--\pageref{lastpage}}
\maketitle


\begin{abstract}
Using photometric data from the Dark Energy Survey and the Wide-field
Infrared Survey Explorer, we estimate photometric redshifts for 105
million galaxies using the nearest-neighbour algorithm.
From such a large data base, 151\,244 clusters of galaxies are
identified in the redshift range of $0.1<z\lesssim 1.5$ based on the
overdensity of the total stellar mass of galaxies within a given
photometric redshift slice, among which 76\,826 clusters are newly
identified and 30\,477 clusters have a redshift $z>1$.
We cross-match these clusters with those in the catalogues identified
from the X-ray surveys and the Sunyaev--Zel'dovich (SZ) effect by the
{\it Planck}, South Pole Telescope and Atacama Cosmology Telescope
surveys, and get the redshifts for 45 X-ray clusters and 56 SZ
clusters. More than 95 percent SZ clusters in the sky region have 
counterparts in our catalogue. We find multiple
optical clusters in the line of sight towards about 15 percent
of SZ clusters.

\end{abstract}

\begin{keywords}
  catalogues --- galaxies: clusters: general --- galaxies: distances and redshifts.
\end{keywords}

\section{Introduction}

As the largest virialized systems in the Universe, clusters of
galaxies trace density peaks in the the large-scale structure.
Galaxy clusters have been identified from the
survey data in optical, X-ray and millimeter bands
\citep[e.g.][]{aco89,pap+11,whl12,bcc+13,planck16b,ggb+19,hsn+21}. Based on
multiband optical photometric observations, clusters of galaxies
can be found through the red sequence of galaxies
\cite[e.g.][]{ogu14,rrb+14} or photometric redshifts of galaxies
\citep[e.g.][]{whl12,yxh+21,zgx+21}. The X-ray observation data show
the X-ray emission from
hot intracluster medium. The Sunyaev--Zel'dovich (SZ) effect 
at the millimeter wavelengths has the advantage to detect massive
clusters up to high redshifts \citep{chr02}. 

The latest data sets observed for the southern Galactic cap provide an
excellent opportunity to uncover clusters at high redshifts. The Dark
Energy Survey \citep[DES;][]{des+16} released photometric data at five
bands for a sky area of 5000 deg$^2$ region down to a limit of $i\sim
24$ magnitude, which is deep enough to detect clusters up to
$z\sim1.5$.  A large number of clusters of galaxies have been
identified from the DES data \citep{rrh+16,abd+21,yxh+21,zgx+21}, but
mostly of $z<1$. In the DES sky region, millimeter survey data have
been obtained by the South Pole Telescope \citep[SPT;][]{caa+11} and
the Atacama Cosmology Telescope \citep[ACT;][]{saa+11}. About 850 and
1800 clusters have been already identified by using the SZ effect from
the SPT and ACT survey data, respectively
\citep{bsd+15,bbs+20,hbs+20,hsn+21}.

In this paper, we combine the DES optical data with the mid-infrared
data from the Wide-field Infrared Survey Explorer
\citep[WISE;][]{wem+10} to identify clusters of galaxies.
In Section 2, we cross-match the DES data with the WISE data for
common galaxies, and estimate their photometric redshifts and stellar
masses. In Section 3, we identify clusters of galaxies from the
estimated photometric redshift data. In Section 4, we compare the
identified clusters with those previously known in the sky region. A
summary is presented in Section 5.

Throughout this paper, we assume a flat Lambda cold dark matter
cosmology taking $H_0=70$ km~s$^{-1}$ Mpc$^{-1}$, $\Omega_m=0.3$, and
$\Omega_{\Lambda}=0.7$.

\section{Photometric redshifts and stellar masses}

Identification of galaxy clusters requires a large number of galaxies
with known redshifts. The photometric redshifts can be used for the
purpose, and can be estimated from photometric data \citep{wh21}. In
addition, stellar masses of galaxies can be derived based on the
mid-infrared luminosity of galaxies, which is useful to identify
density peaks.

\subsection{Photometric data of galaxies}

The DES\footnote{https://www.darkenergysurvey.org/} made photometric
observations at five broad optical bands ($grizy$), covering the
southern Galactic cap of $\sim$5000 deg$^2$ \citep{des+16}. The survey
reaches a 10$\sigma$ depth of $i\sim 24$ (AB magnitude) for point
sources. The median full width at half-maximum of the
point spread function is 0.88 arcsec in the $i$-band.  From the latest
public DES data release 2 \citep[DR2,][]{desdr2}, we get 399 million
galaxies detected above 5$\sigma$ in at least one band and with the
magnitudes at least in the $r$, $i$ and $z$ bands. The galaxies are
selected with the flags of IMAFLAGS$_{\rm ISO}=0$, FLAGS$<4$ and an
extended morphology classification of SPREAD\_MODEL$\ge2$.

The WISE observed the whole sky at four mid-infrared bands of ${\rm
  W1}$, ${\rm W2}$, ${\rm W3}$, and ${\rm W4}$ \citep{wem+10}. The
angular resolutions of the WISE are 6.1 and 6.5 arcsec in the ${\rm
  W1}$ and ${\rm W2}$ bands, respectively. We use the ${\rm W1}$ and
${\rm W2}$-band data from the five-year coadd image
data\footnote{http://unwise.me/}, named as unWISE, which is deep
enough to detect massive galaxies up to a redshift of $z\sim2$
\citep{lang14,smg19}. The unWISE catalogue contains 208 million
sources with ${\rm W1}$-band data available in the sky coverage of the
DES DR2.

The spread function of the unWISE data is much wider than that of the
DES data, so that very close DES galaxies may not be resolved in the
unWISE. As in \citet{wh21}, we perform one-to-one match between the
DES galaxies and the unWISE sources within a radius of 2 arcsec. The
closest source within this radius is adopted for the targeted
object. The such matched 105 million objects are named as
DES$\times$unWISE galaxies in this paper.

\subsection{Photometric redshifts of galaxies}
\label{photozg}

Photometric redshifts of DES$\times$unWISE galaxies are determined
based on the galaxy colours and their nearest-neighbour spectroscopic
redshifts in a training sample, following \citet{wh21}.

\begin{figure}
\centering \includegraphics[width = 0.4\textwidth]{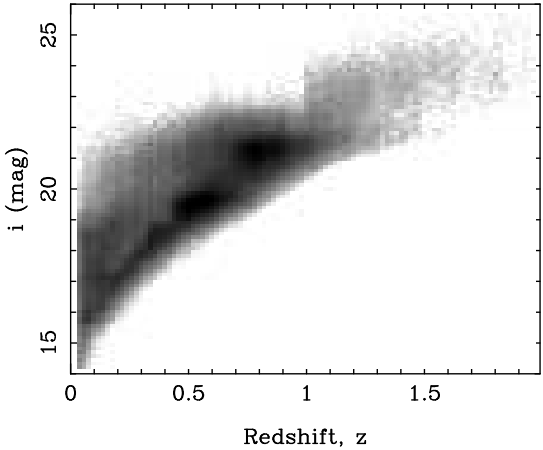}
\caption{The training sample has a wide distribution in the redshift and
is deep enough in the $i$-band.}
\label{trainingmag}
\end{figure}

\begin{figure}
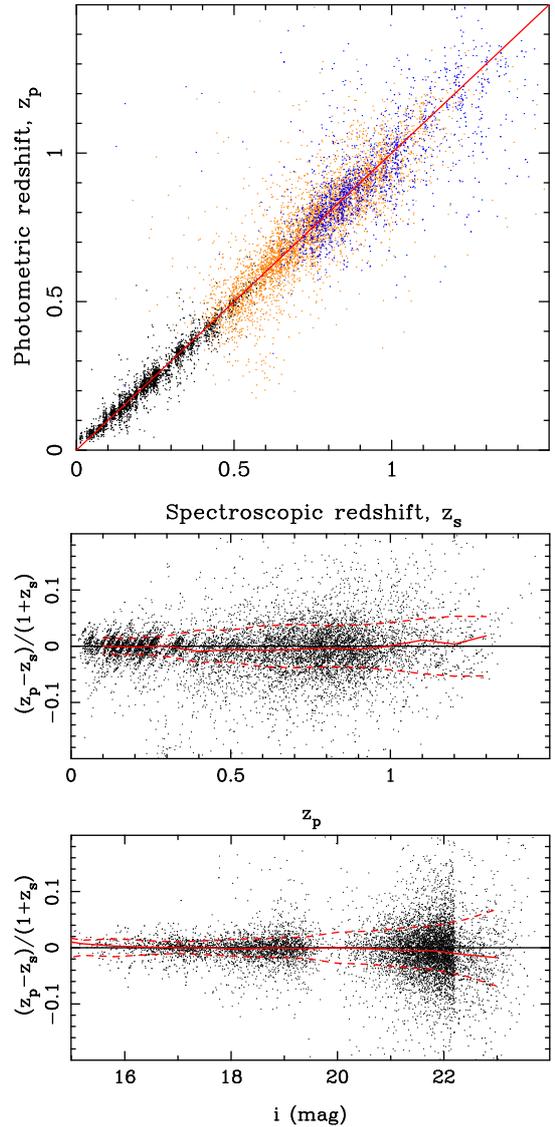

  \centering \includegraphics[width = 0.4\textwidth]{f2a.eps}
  \centering \includegraphics[width = 0.4\textwidth]{f2b.eps}
  \centering \includegraphics[width = 0.4\textwidth]{f2c.eps}
\caption{Upper panel: verification of estimated redshifts by comparing
  them with spectroscopic redshifts of galaxies in the GAMA (black),
  VIPERS DR2 (qFlag$=$2--3, orange) and DEEP2 (blue) data. A great
  consistence is shown in a very wide range of redshift, though the
  dispersion of data becomes larger at higher redshifts. Middle panel:
  the difference between estimated photometric redshifts and
  spectroscopic redshifts plotted against photometric redshift.
  Bottom panel: the difference plotted against $i$-band magnitude. The
  dashed lines indicate $\pm \sigma_{\Delta z}$ and the solid line
  indicates the bias.}
\label{biascheck}
\end{figure}

\begin{figure}
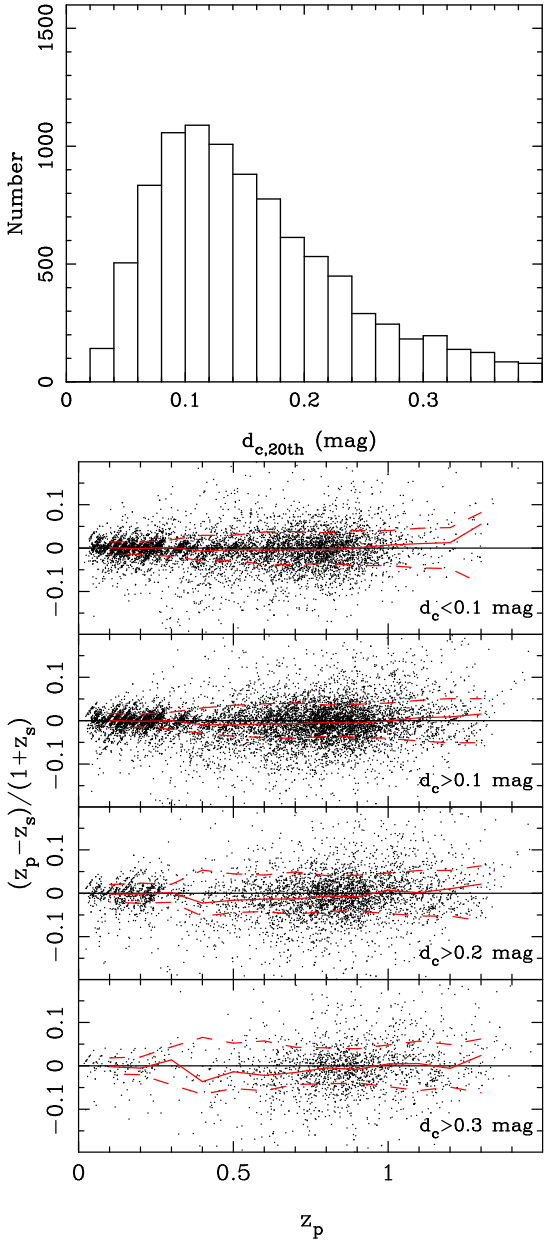

\centering \includegraphics[width = 0.4\textwidth]{f3a.eps}
\centering \includegraphics[width = 0.4\textwidth]{f3b.eps}
\caption{Upper: distribution of total colour offset of the 20th neighbour
  in the multidimensional colour space. Lower: redshift differences for four
  redshift estimates by using only the galaxies among the 20 nearest neighbours
  with colour offsets of $d_c<0.1$ mag, or $d_c>0.1$ mag, or $d_c>0.2$ mag or
  $d_c>0.3$ mag, respectively. The dashed lines indicate $\pm \sigma_{\Delta z}$
  and the solid line indicates the bias.}
\label{biascheck2}
\end{figure}

\begin{figure}
\centering \includegraphics[width = 0.4\textwidth]{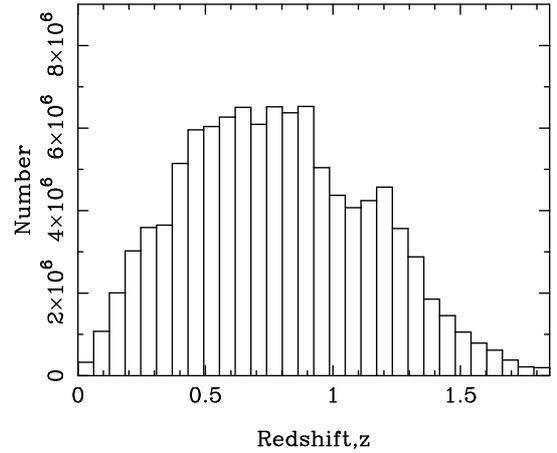}
\caption{Number distribution of photometric redshifts of 
  DES$\times$unWISE galaxies.}
\label{histgz}
\end{figure}

The training sample of spectroscopic redshifts are obtained from the
Sloan Digital Sky Survey (SDSS) data release 16 \citep{sdssdr16},
PRIMUS DR1 \citep{cmb+13}, VIPERS PDR2 \citep{sgg+18}, VVDS
\citep{lcc+13}, and 3D-HST \citep{mbv+16}.  We adopt the same redshift
flags as in \citet{wh21}, and get spectroscopic redshifts for 460\,653
galaxies, among which 335\,896 galaxies have DES$\times$unWISE
magnitudes in at least six bands of $grizyW1$. To have a good
redshift coverage for faint galaxies, we also include 6341
galaxies from the Hyper Suprime-Cam Subaru Strategic Program
(HSC-SSP)$\times$unWISE data \citep{wh21} into the training sample,
which have spectroscopic redshifts in the range of $1<z<2$. For these
galaxies, the HSC-SSP $grizy$ magnitudes are converted to the DES
$grizy$ magnitudes following conversion equations in
\citet{desdr2}. In total, we have 342\,237 galaxies from
the DES$\times$unWISE and HSC-SSP$\times$unWISE for the training sample.
Fig.~\ref{trainingmag} shows the wide distribution of galaxies in
the redshift-magnitude diagram. The faint end of the training sample
extends to $i>24$ and the redshift extends to $z>1.5$,
suggesting that the training sample can be used to faint
galaxies near the limit of the DES$\times$unWISE galaxies.

For each of the DES$\times$unWISE galaxies, we estimate the
photometric redshift as being the median of the spectroscopic
redshifts of 20 nearest neighbours in multidimensional colour space
($g-r$, $r-i$, $i-z$, $i-y$, $i-{\rm W1}$, and ${\rm W1}-{\rm W2}$) of
the training sample \citep{dss16,wh21}. The uncertainty of each
photometric redshift ${\rm \sigma_{zp}}$ is estimated as being the
standard deviation of the spectroscopic redshifts of these 20 nearest
neighbours. If an object has many neighbours from the training objects
in the colour space with a similar redshift, the photometric redshift
would have a small uncertainty; if a galaxy is located in a sparse
region in the colour space where less objects are available, the
photometric redshift would have to be estimated from the largely
scattered spectroscopic redshifts and thus has a large uncertainty.
Therefore the uncertainty of each photometric redshift indicates the
reliability of the estimated photometric redshift.

We verify the accuracy of such estimated redshifts by comparing them
with spectroscopic redshifts of galaxies by using an independent
testing sample from the GAMA DR2 \citep{lbd+15}, VIPERS PDR2
\citep{sgg+18}, and DEEP2 DR4 \citep{mnc+13}. Most galaxies in the GAMA
data have a low redshift of $z<0.5$. The VIPERS PDR2 data have a high
($>99$ percent) confidence on redshift for the galaxies with qFlag$=$3--4
and have been included in the training sample. Here, we adopt the
VIPERS redshifts for the galaxies at $0.5<z<1.2$ with qFlag$=$2--3,
which have a slightly lower but also excellent ($>95$ percent)
confidence. The DEEP2 galaxies mostly have redshifts in the range of
$0.7<z<1.4$. These data are magnitude-limited without any colour
selection. We discard the common galaxies with those in the training
sample and obtain a testing sample of 9857 galaxies at
$z\lesssim1.4$. Their photometric redshifts are estimated and compared
with the spectroscopic redshifts, as shown in
Fig.~\ref{biascheck}. The redshift difference, $(z_p-z_s)/(1+z_s)$, is
plotted against the photometric redshift (middle panel). The
systematic bias, defined as the median value of the difference, is
generally small at $z<1$ and the largest bias is about $-0.008$ at
$z\sim0.5$.  The bias is about 0.01 at $z>1.1$. The redshift
uncertainty, defined as $\sigma_{\Delta z}=1.48\times {\rm
  median}(|z_p-z_s|/(1+z_s))$, generally increases with redshift from
0.013 at $z\sim0.1$ to 0.052 at $z\sim1.3$. The redshift difference
also slightly depends on $i$-band magnitude (bottom panel of
Fig.~\ref{biascheck}). Naturally the galaxies at the faint end have the largest
bias, which is about $-0.018$ at $i=23$.

In our algorithm for photometric redshifts of galaxies, we fix
the number of 20 nearest neighbours to estimate photometric
redshifts. Some of the neighbours may have a large colour offset from
the targeted galaxy in the colour space, especially at high redshifts
with sparse training samples. We test if such neighbours with large
offsets cause a significant bias for the estimated photometric
redshift. In the upper panel of Fig.~\ref{biascheck2}, we show the
total colour offset distribution of the 20th nearest neighbour in the
multidimensional colour space of $g-r$, $r-i$, $i-z$, $i-y$, $i-{\rm
  W1}$, and ${\rm W1}-{\rm W2}$ for the 9857 testing galaxies. The
distribution has a median value of 0.14 mag, and 12 per cent of galaxies have
offsets larger than 0.3 mag. In the lower panels of
Fig.~\ref{biascheck2}, we calculate four testing photometric redshifts
by using the training galaxies among the 20 nearest neighbours with
colour offsets $d_c<0.1$ mag, or only galaxies with $d_c>0.1$ mag, or
$>0.2$ mag or $>0.3$ mag, respectively. We compare the redshift
differences and find that the biases are not significant except for
the cases with small number of galaxies.

Using the training sample of 342\,237 galaxies, we then estimate the
photometric redshifts for 105 million DES$\times$unWISE galaxies. The
obtained photometric redshifts have a wide distribution up to
$z\sim1.7$ with a peak at $z\sim0.7$ (see Fig.~\ref{histgz}), which
enable us to identify a large sample of clusters to redshifts $z>1$.

\begin{figure}
\centering \includegraphics[width = 0.4\textwidth]{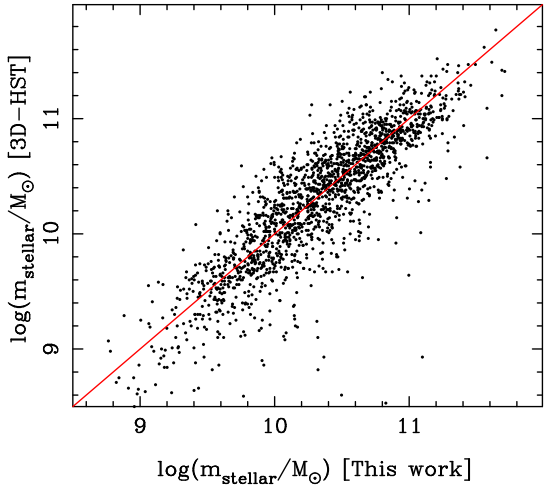}
\caption{Comparison of our stellar masses estimated from ${\rm
    W1}$-band luminosity with the accurate values obtained from the
  3D-HST survey shows a good agreement. The solid line indicates the
  equality.}
\label{3dhstmass}
\end{figure}

\subsection{Stellar mass of galaxies}

With photometric redshifts estimated, the stellar masses, $m_{\rm
  stellar}$, can be derived for the 105 million galaxies in the
DES$\times$unWISE catalogue from the ${\rm W1}$-band luminosity
following the procedures of our previous work \citep{wh21}. Comparison
of our estimates with the accurate values obtained from the 3D-HST
survey \citep{mbv+16}, as shown in Fig.~\ref{3dhstmass}, shows that
they are in good agreement, with a scatter of only 0.23 dex.

\begin{figure}
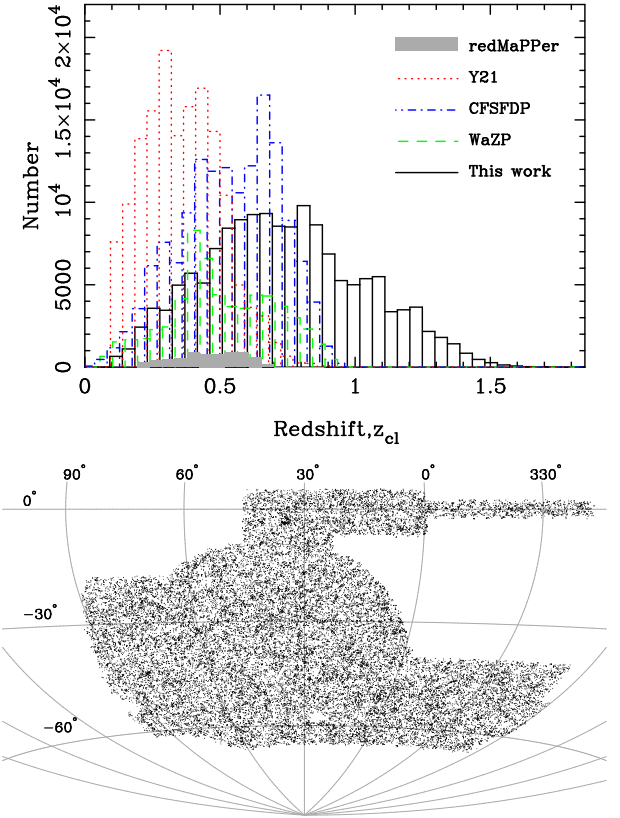

  \centering
  \includegraphics[width = 0.42\textwidth]{f6a.eps}
 \\[2mm] 
  \includegraphics[width = 0.45\textwidth]{f6b.eps}
\caption{{\it Upper panel:} The redshift distribution of 151\,244
  clusters in our catalogue compared with those in the DES footprint,
  including clusters from the catalogues of redMaPPer \citep{rrh+16},
  Y21 \citep[only clusters with more than 10
    members in][]{yxh+21}, CFSFDP \citep{zgx+21}, WaZP \citep{abd+21}. {\it
    Lower panel:} The sky distribution of newly identified high
  redshift clusters of $z_{\rm cl}>1.0$.}
\label{histcz}
\end{figure}

\begin{table*}
\begin{minipage}{160mm}
\caption[]{The 151\,244 clusters of galaxies identified from the
  DES$\times$unWISE data.}
\begin{center}
\setlength{\tabcolsep}{1mm}
\begin{tabular}{crrrcccccrccc}
\hline
\mc{1}{c}{Cluster ID}& \mc{1}{c}{Name}&\mc{1}{c}{R.A.} & \mc{1}{c}{Dec.} & \mc{1}{c}{$z_{\rm cl}$} & \mc{1}{c}{$i_{\rm BCG}$} &
\mc{1}{c}{${\rm W1_{BCG}}$} &  \mc{1}{c}{SNR} & \mc{1}{c}{$r_{500}$} &
\mc{1}{c}{$\lambda_{500}$} & \mc{1}{c}{$M_{500}$} &\mc{1}{c}{$N_{\rm gal}$} & \mc{1}{c}{Other catalogues} \\
\mc{1}{c}{(1)} & \mc{1}{c}{(2)} & \mc{1}{c}{(3)} & \mc{1}{c}{(4)} & \mc{1}{c}{(5)} & 
\mc{1}{c}{(6)} & \mc{1}{c}{(7)} & \mc{1}{c}{(8)} & \mc{1}{c}{(9)} & \mc{1}{c}{(10)} &
\mc{1}{c}{(11)} & \mc{1}{c}{(12)} & \mc{1}{c}{(13)} \\
\hline
  1& WH J000000.5$+$021911 & 0.00200 & $  2.31979$ & 0.4192& 18.738& 17.964 &  6.75& 0.549 & 18.77 &  0.86 & 15& WH21     \\
  2& WH J000001.3$-$640959 & 0.00555 & $-64.16639$ & 0.6423& 18.565& 17.360 & 10.17& 0.673 & 36.48 &  1.62 & 12&          \\
  3& WH J000001.4$-$521956 & 0.00563 & $-52.33236$ & 0.5829& 19.249& 18.207 &  6.00& 0.590 & 21.61 &  0.98 & 12& redMaPPer\\
  4& WH J000002.3$-$475113 & 0.00943 & $-47.85358$ & 0.7734& 20.239& 18.481 &  5.48& 0.487 & 17.47 &  0.80 & 13& CFSFDP   \\
  5& WH J000002.8$-$474415 & 0.01186 & $-47.73740$ & 0.3169& 17.305& 17.307 &  7.53& 0.548 & 17.18 &  0.79 & 11&          \\
  6& WH J000003.1$-$033245 & 0.01274 & $ -3.54574$ & 0.6056& 18.754& 17.483 & 14.00& 0.722 & 43.60 &  1.92 & 22& WHL      \\
  7& WH J000003.8$-$010154 & 0.01592 & $ -1.03153$ & 0.7327& 19.759& 18.242 &  6.06& 0.621 & 26.65 &  1.20 & 10& WaZP     \\
  8& WH J000003.9$-$525115 & 0.01610 & $-52.85407$ & 0.7657& 20.561& 18.513 &  6.34& 0.559 & 20.93 &  0.95 & 12&          \\
  9& WH J000004.2$-$393257 & 0.01733 & $-39.54926$ & 0.6064& 19.742& 18.163 &  6.69& 0.542 & 20.50 &  0.93 & 13& CFSFDP   \\
 10& WH J000004.2$+$021941 & 0.01742 & $  2.32799$ & 0.6228& 19.247& 17.721 &  6.47& 0.541 & 20.33 &  0.92 &  9& WHL      \\
\hline
\end{tabular}
\end{center}
{Note.
Column 1: Cluster ID;
Column 2: Cluster name with J2000 coordinates of cluster; 
Columns 3 and 4: Right Ascension (R.A. J2000) and Declination (Dec. J2000) of cluster BCG (in degree);
Column 5: cluster redshift $z_{\rm cl}$; 
Columns 6--7: BCG magnitudes (AB system) in $i$ and ${\rm W1}$ bands, respectively;
Column 8: SNR for cluster detection;
Column 9: cluster radius, $r_{500}$, in Mpc; 
Column 10: cluster richness;
Column 11: derived cluster mass, in units of $10^{14}~M_{\odot}$;
Column 12: number of member galaxy candidates within $r_{500}$; 
Column 13: Reference notes for previously known clusters: WHL \citep{whl12,wh15}, SPT \citep{bsd+15,bbs+20,hbs+20},
redMaPPer \citep{rrh+16}, WHY18 \citep{why18}, RASS \citep{kgm+19}, ACT \citep{hsn+21},
WH21 \citep{wh21}, Y21 \citep{yxh+21}, CFSFDP \citep{zgx+21}, WaZP \citep{abd+21}.\\
(This table is available in its entirety in a machine-readable form.)
}
\label{tab1}
\end{minipage}
\end{table*}

\section{Clusters identified from the DES$\times$unWISE galaxies}

Following \citet{wh21}, we identify a cluster of galaxies based on the
overdensity of galaxy stellar mass within a photometric redshift slice
around each massive galaxy of $m_{\rm stellar}\ge5\times10^{10}~M_{\odot}$
that is temporally assumed to be the brightest cluster galaxy (BCG) of a cluster
candidate.
The redshift of a cluster candidate, $z$ (or $z_{\rm cl}$ for a
finally identified cluster), is taken to be the median value of the
photometric redshifts of member galaxy candidates within a photometric
redshift slice of $z\pm\Delta z$ and a projected radius of
$r_1$. The half of the slice thickness is taken as being
\begin{equation}
  \Delta z=\left\{
\begin{array}{ll}
   0.04\,(1+z)           &     \mbox{for $z\leq 0.7$}\\
   0.15\,z-0.037       &     \mbox{for $z> 0.7$}
\end{array}
\right. \,,
\label{pzslice}
\end{equation}
and the projected radius is taken as 
\begin{equation}
r_1=1.0\,E(z)^{-2/3}~{\rm Mpc},
\end{equation}
where $E(z)=\sqrt{\Omega_{\Lambda}+\Omega_m(1+z)^3}$ (there is a minor mistake for the
equation of $r_1$ in Wen \& Han 2021). 
Member galaxy candidates are excluded if their photometric redshifts 
are unreliable with ${\rm \sigma_{zp}}>2\,\Delta z$. From these member galaxy
candidates, we get the total stellar mass, $m_{\rm stellar,r1}$ within
the radius $r_1$. The cluster radius, $r_{500}$, within which the mean
density of a cluster is 500 times of the critical density of the
universe, is estimated from a calibrated $r_{500}$--$m_{\rm  stellar,r1}$
relation based on a sample of X-ray clusters
\citep{wh21}. Then, the total stellar mass within the estimated radius
$r_{500}$ is calculated from the galaxy data, which is later defined as the
richness of a cluster, $\lambda_{500}$, as in \citet{wh21}. The
overdensity is expressed as the detection signal-to-noise ratio, SNR,
which is calculated based on the overdensity of galaxy stellar mass
within a projected radius of 0.5 Mpc related to the density
fluctuations in the surrounding regions of the same redshift slice. A
cluster is identified with the threshold of ${\rm SNR}\ge5$ and a
richness of $\lambda_{500}\ge15$, and also the number of member galaxy
candidates within $r_{500}$ as being $N_{\rm gal}\ge6$. Finally, the repeated 
entries within a photometric redshift slice of $z\pm1.5\Delta
z$ and a projected distance of $1.5\,r_{500}$ are cleaned so that only
one cluster candidate with the largest richness is adopted. The low-redshift
clusters of $z_{\rm cl}<0.1$ are discarded because some
bright BCGs have the flag of IMAFLAGS\_ISO$\neq$0 and are not included
in the DES$\times$unWISE data.

From the DES$\times$unWISE galaxies with estimated redshifts, we find
151\,244 clusters as listed in Table~\ref{tab1}, among which 76\,826
clusters are newly identified and 74\,418 clusters are previously known
\citep{bsd+15,bbs+20,wh15,wh21,rrh+16,why18,kgm+19,hbs+20,abd+21,hsn+21,yxh+21,zgx+21}.
Cluster masses are derived by using the mass--richness relation in \citet{wh21}.
The 2\,454\,302 member galaxy candidates within $r_{500}$ for these clusters
are also publicly available on
web-page\footnote{http://zmtt.bao.ac.cn/galaxy\_clusters/}.
Fig.~\ref{histcz} shows the redshift distribution of identified
clusters in the range of $0.1<z_{\rm cl}\lesssim1.5$ with a wide peak
at about $z_{\rm cl}\sim0.7$. Much more clusters are found at
higher redshifts than previous cluster findings. We get 30\,477
clusters with a redshift of $z_{\rm cl}>1$, and they are widely
distributed in the sky (see low panel of Fig.~\ref{histcz}).

\begin{figure}
\centering \includegraphics[width = 0.4\textwidth]{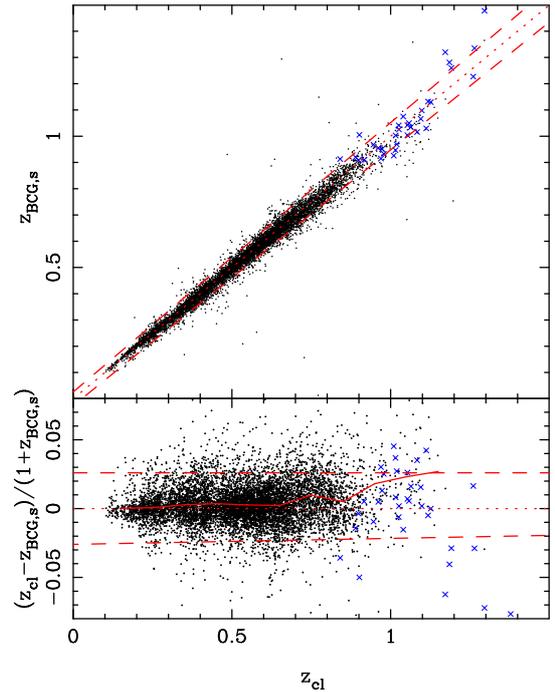}
\caption{Comparison between the spectroscopic redshifts of BCGs with
  cluster photometric redshifts. The dashed lines indicate the
  2$\sigma$ uncertainty. Blue crosses are clusters with known
  spectroscopic redshifts we obtained from the NED.}
\label{spzc}
\end{figure}

To evaluate the accuracy of cluster photometric redshift, in
Fig.~\ref{spzc}, we compare the estimated cluster redshifts with
available spectroscopic redshifts for their BCGs of 8739 clusters plus
the spectroscopic redshifts for 32 clusters of $z>0.9$ from the
NASA/IPAC Extragalactic Database
(NED)\footnote{http://ned.ipac.caltech.edu/}. The uncertainty of
cluster redshift, defined as $1.48\times {\rm median}(|z_{\rm
  cl}-z_{\rm BCG,s}|/(1+z_{\rm BCG,s}))$, is about 0.013. Only 1.6 per cent
of our clusters have photometric redshifts that deviate from
spectroscopic redshifts larger than $0.05(1+z)$. A few outliers may be
a false identification in the case that foreground or background
galaxies are miss taken as BCGs. As shown in Fig.~\ref{biascheck}, the
uncertainty of galaxy photometric redshift rapidly increases at
$z>0.9$, making that the estimated cluster redshift has a larger
uncertainty and a systematic bias of about 0.02 at $z_{\rm cl}\sim
1$. Due to the small number of clusters with spectroscopic redshift
measurements, the uncertainty of cluster redshifts is hard to assess
at $z>1$.

\begin{figure*}
\resizebox{58mm}{!}{\includegraphics{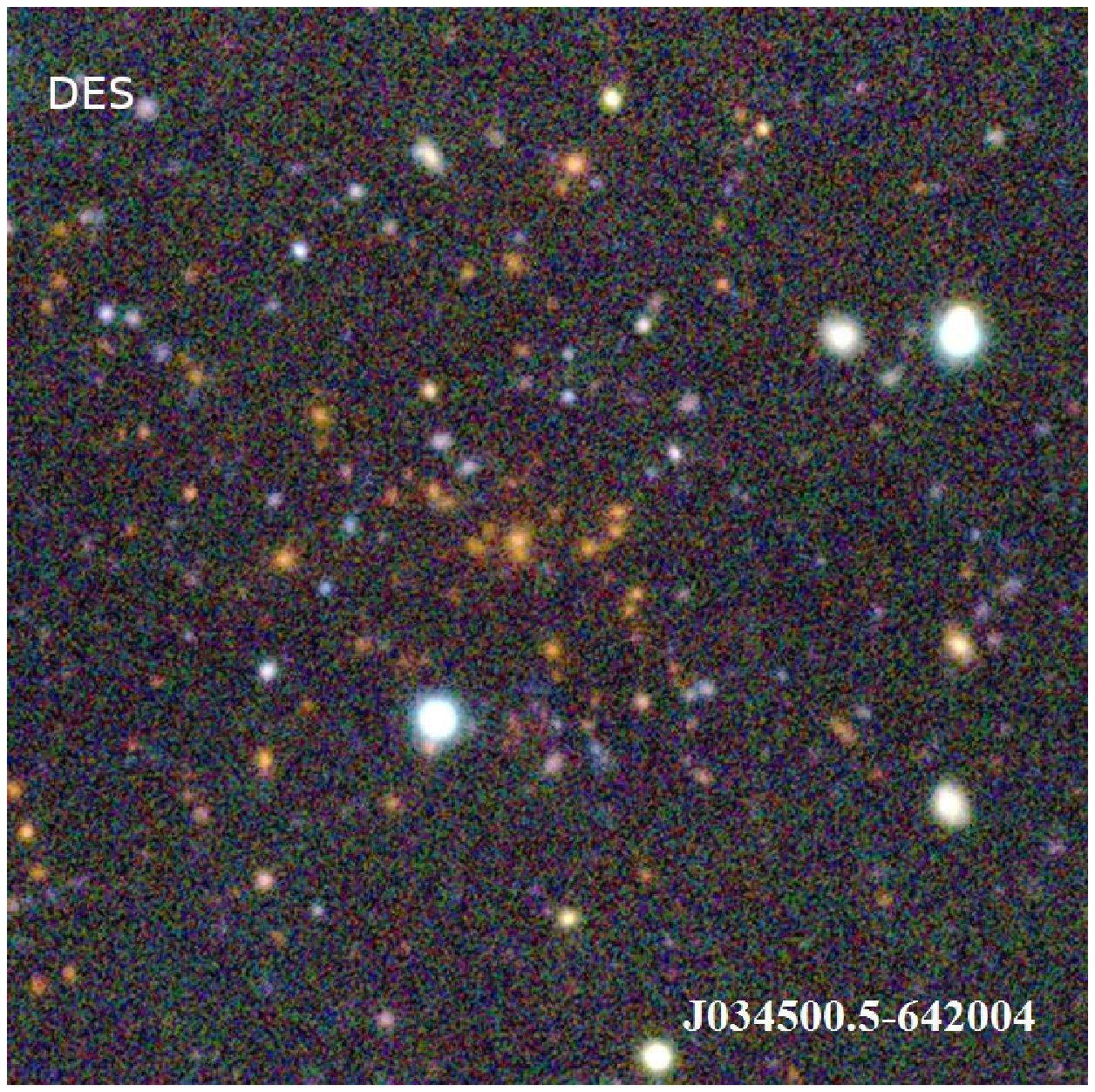}}
\resizebox{58mm}{!}{\includegraphics{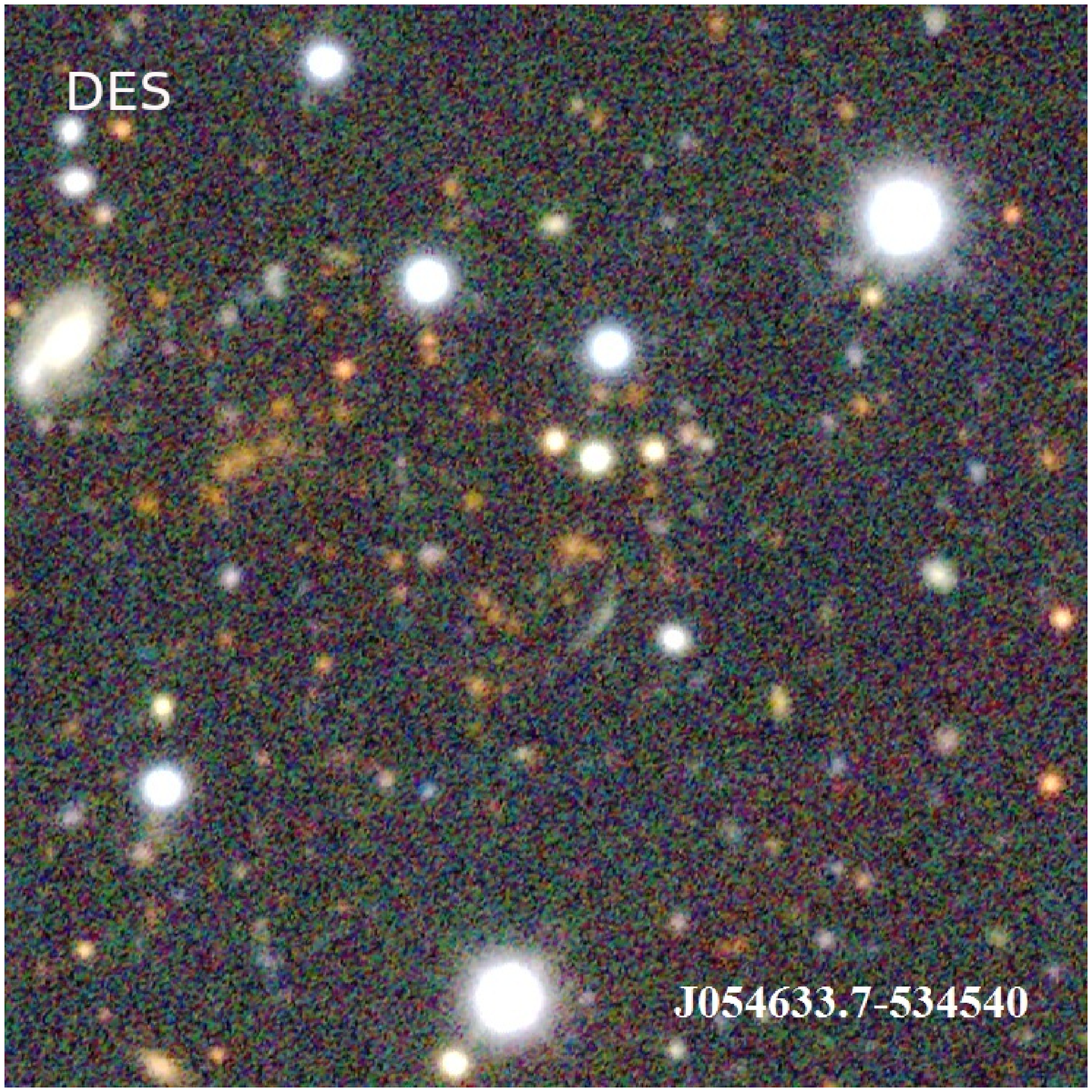}}
\resizebox{58mm}{!}{\includegraphics{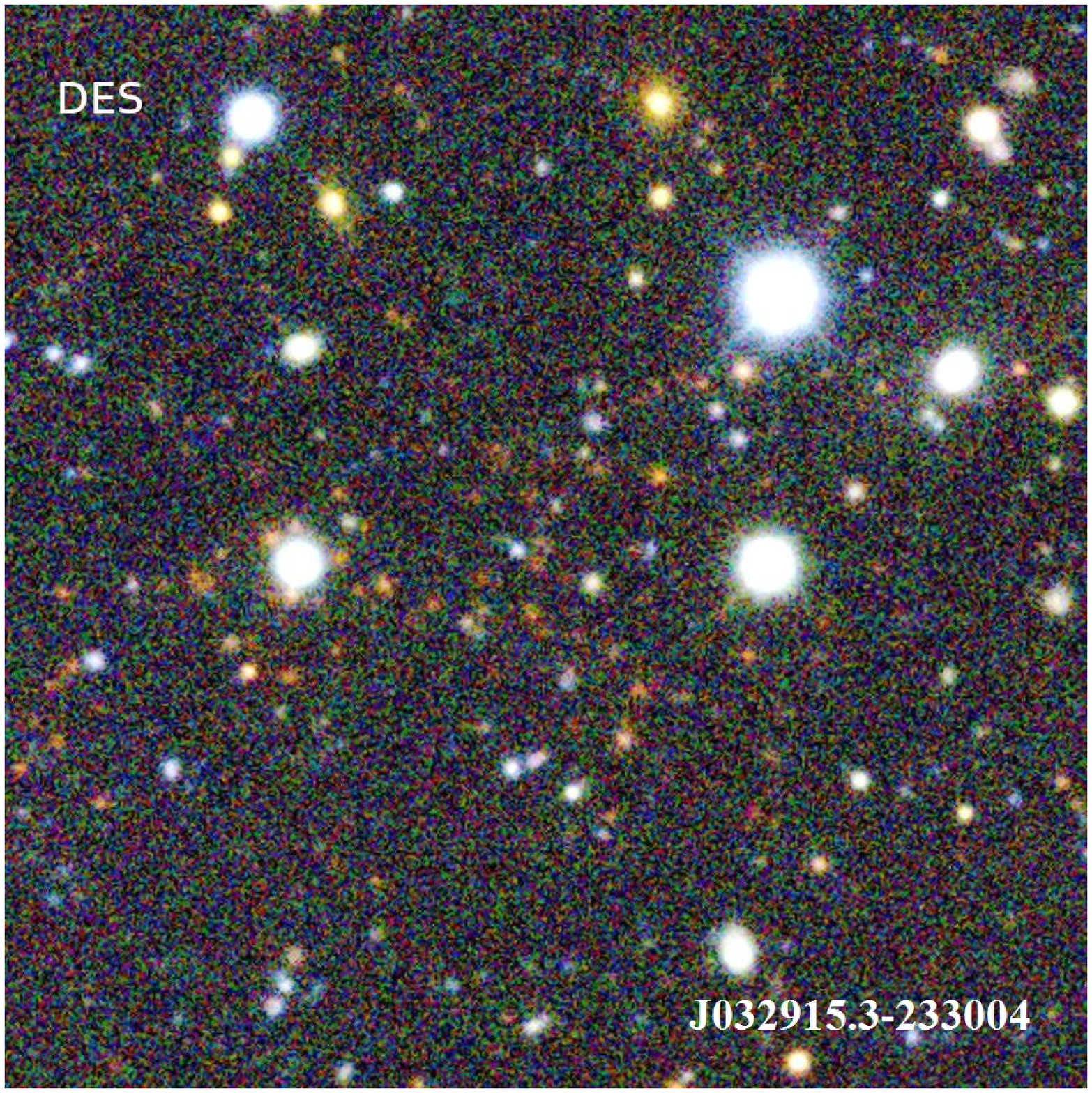}}
\caption{DES composite ($riz$) colour images of three
  identified clusters at redshifts of $z_{\rm cl}=1.0055$ (left),
  1.0959 (middle), and 1.2602 (right) and with
  richnesses of $\lambda_{500}=106.19$, 126.85, and 139.60,
  respectively. The images have a size of 2 arcmin $\times$ 2 arcmin.}
\label{example}
\end{figure*}

\begin{figure*}
\resizebox{58mm}{!}{\includegraphics{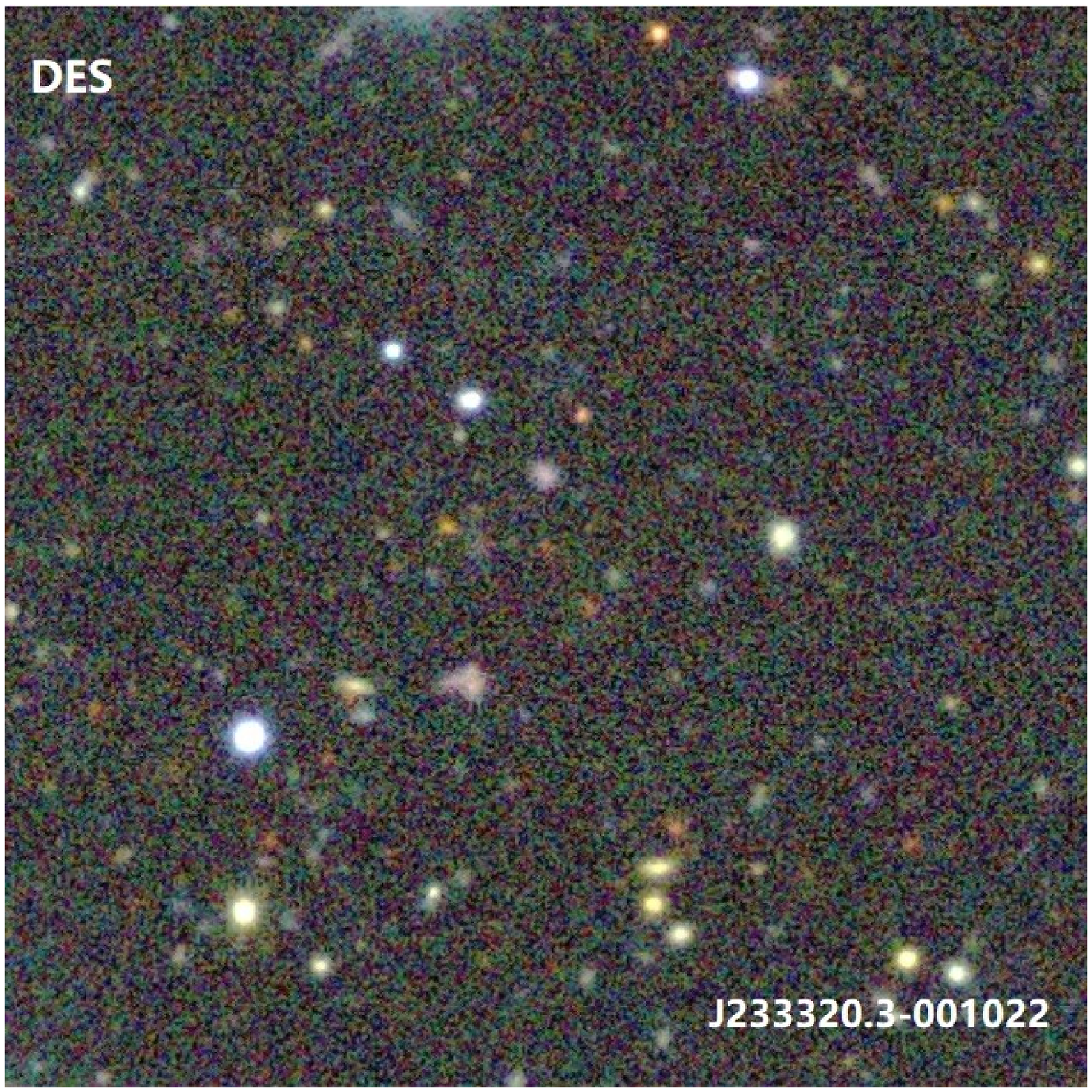}}
\resizebox{58mm}{!}{\includegraphics{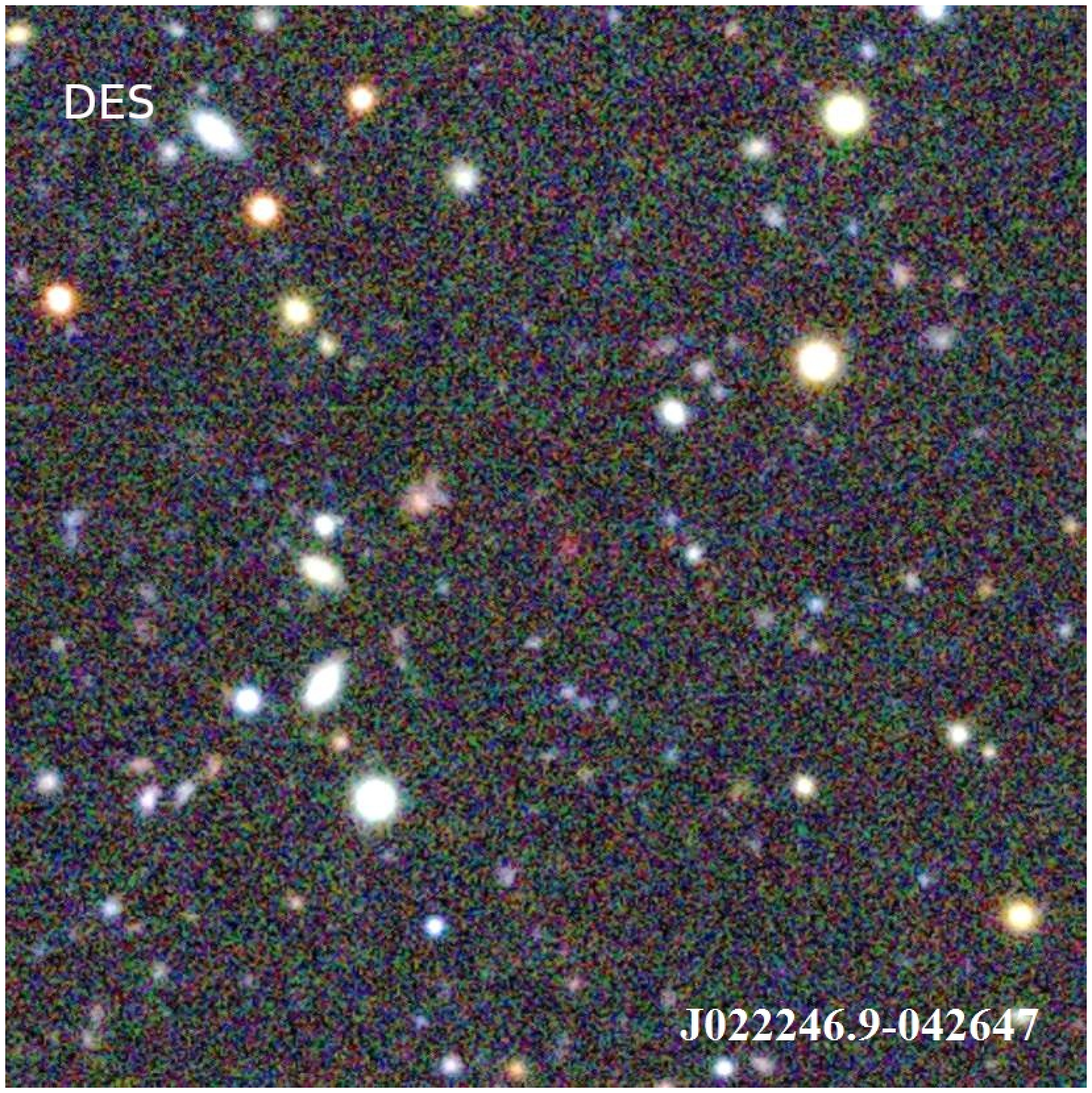}}
\resizebox{58mm}{!}{\includegraphics{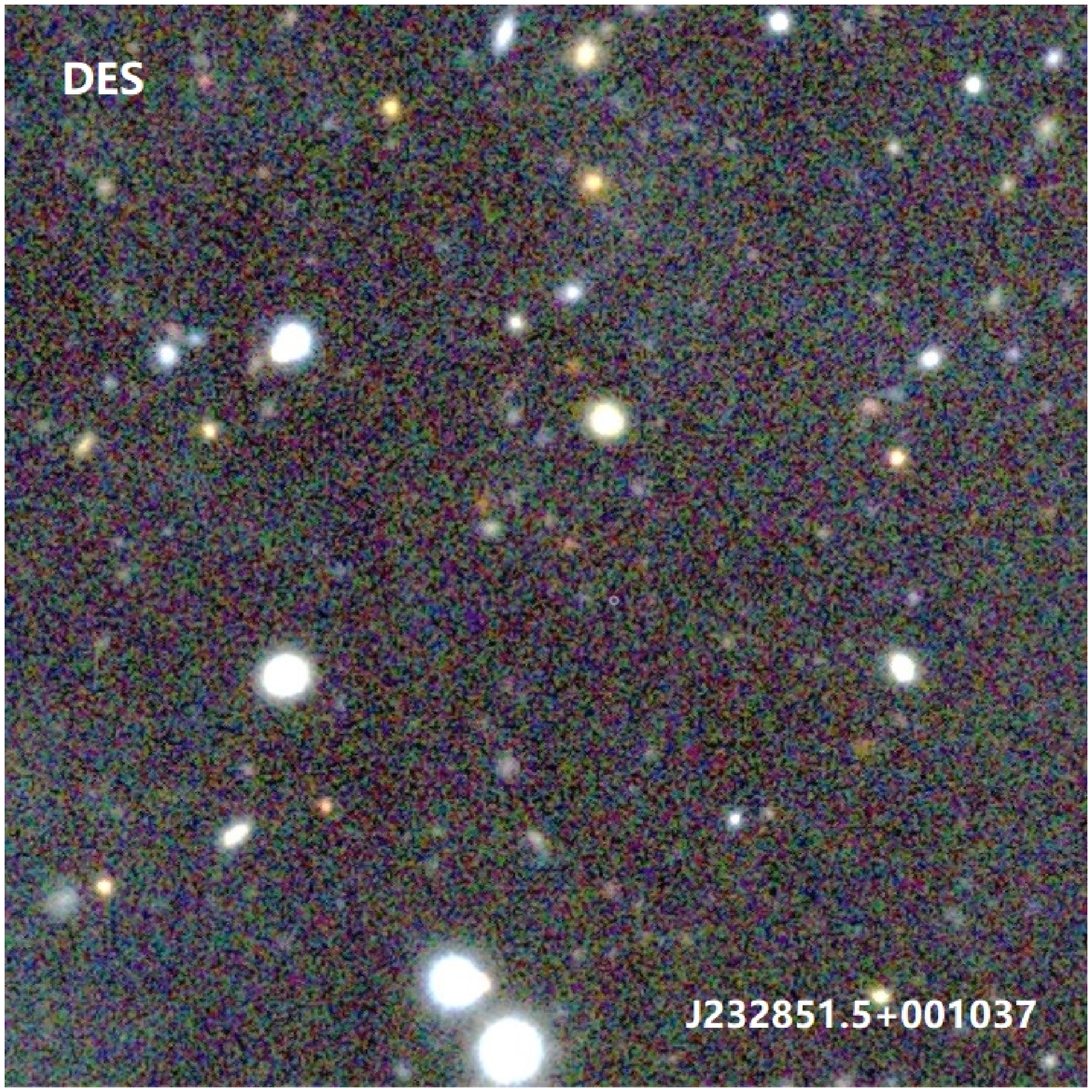}}
\resizebox{58mm}{!}{\includegraphics{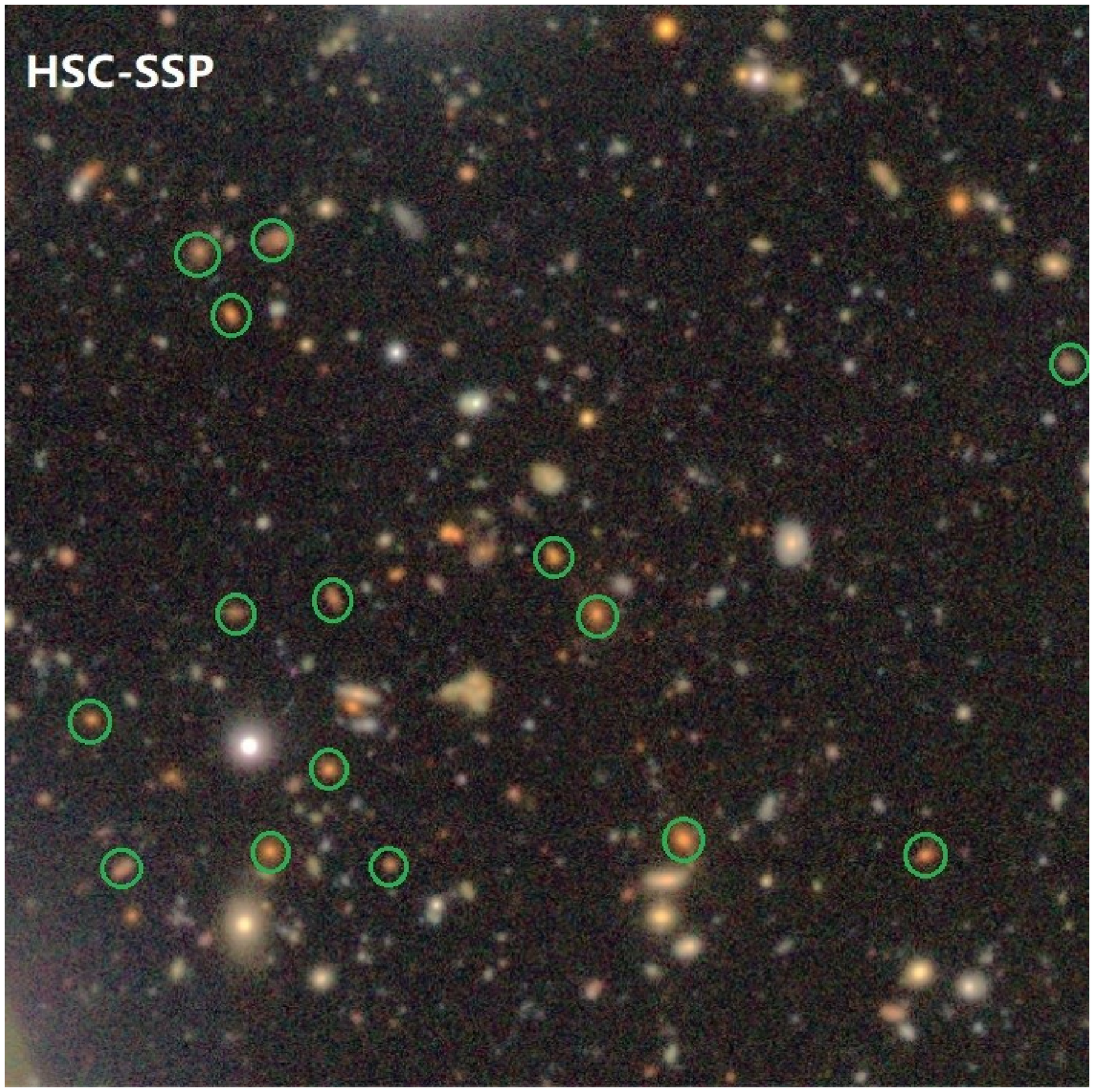}}
\resizebox{58mm}{!}{\includegraphics{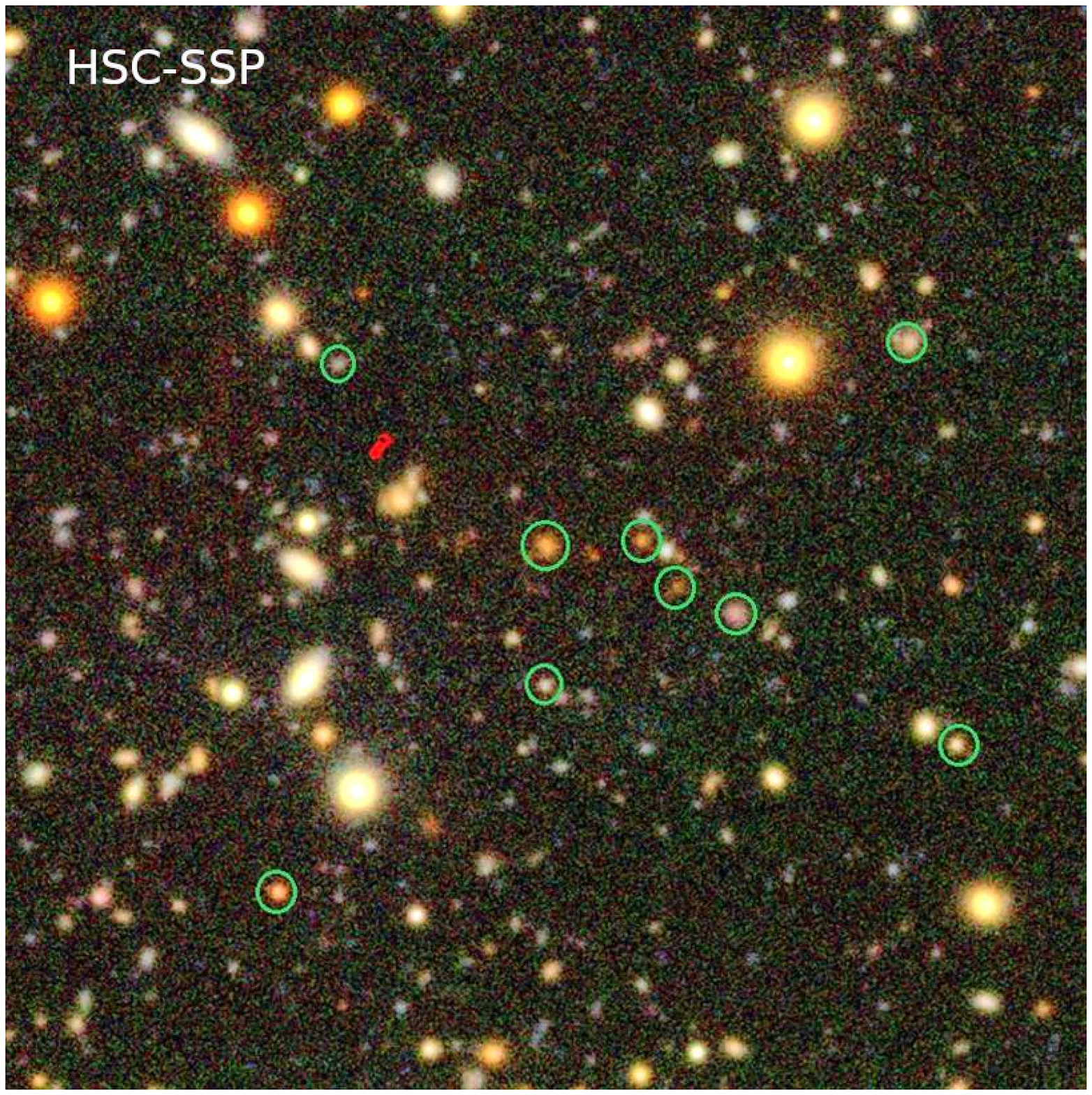}}
\resizebox{58mm}{!}{\includegraphics{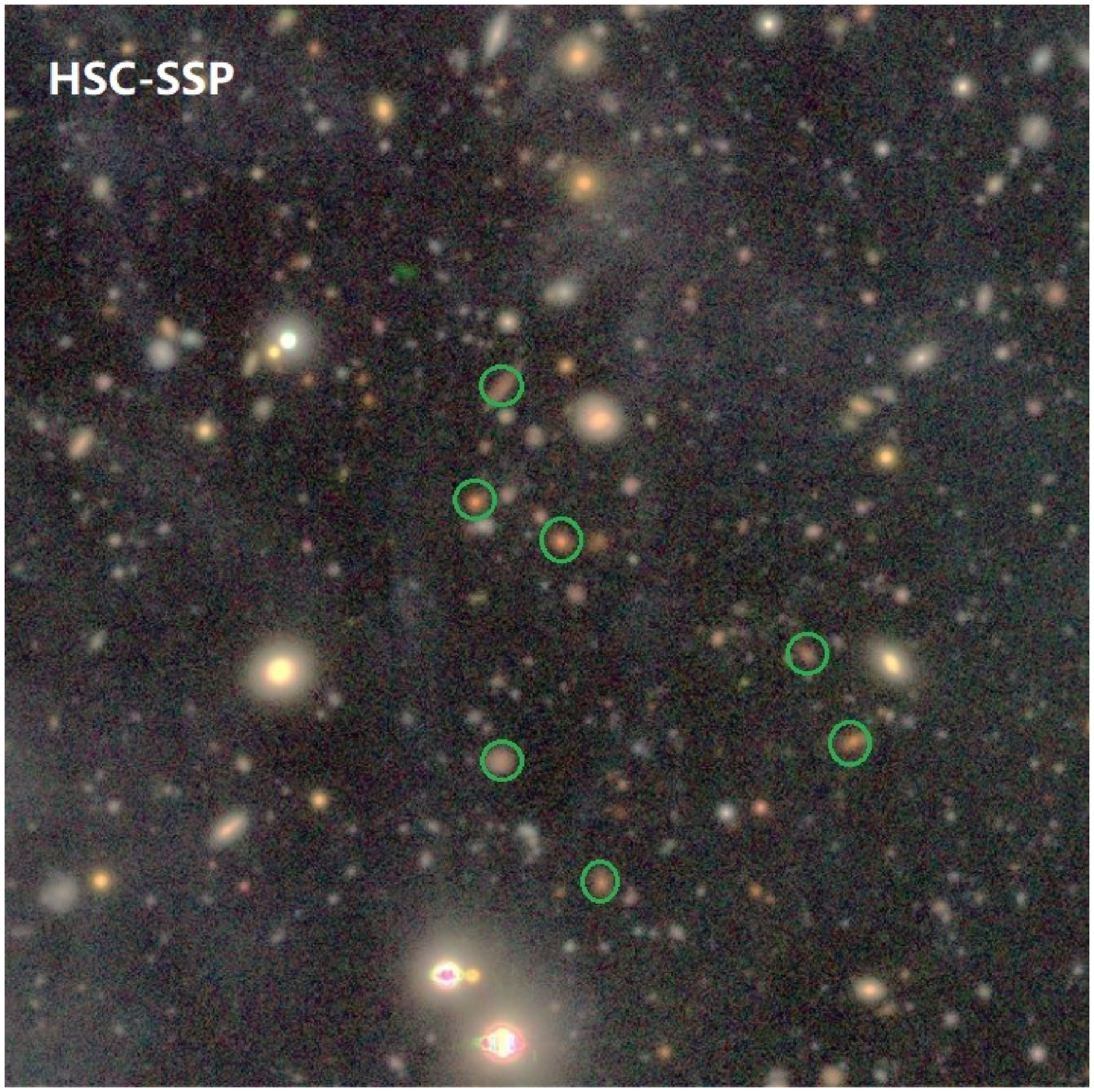}}
\caption{Upper panels:
  DES colour images for three clusters at higher redshifts of $z_{\rm
    cl}=1.3046$ (left), 1.3510 (middle), and 1.3957 (right)
   with richnesses of $\lambda_{500}=42.03$, 27.26,
  and 20.77, respectively. Lower panels: same clusters as in the upper
  panels but the HSC-SSP Deep colour images. The green circles
  indicate the member galaxy candidates of the clusters.}
\label{example2}
\end{figure*}

We visually inspect the colour images of the identified clusters from
the DES$\times$unWISE data for a verification of our catalogue,
especially at high redshifts. Fig.~\ref{example} shows the DES colour
($riz$) images for three rich clusters at redshifts of $z_{\rm
  cl}=1.0055$, 1.0959, and 1.2602, respectively. The clusters
have a richness of $\lambda_{500}=106.19$, 126.85, and 139.60,
respectively, and show many red galaxies concentrated. To check the
clusters at higher redshifts, we inspect the colour images of both DES
and HSC-SSP Deep data for the clusters in a small overlapped region.
Fig.~\ref{example2} also shows the colour images
for three high-redshift clusters at $z_{\rm cl}=1.3046$, 1.3510, and 1.3957,
respectively, and with a richness of $\lambda_{500}=42.03$, 27.26, and
20.77, respectively. Only a few massive
galaxies are selected as member galaxy candidates. They are very
indistinct in the DES images, but clearly shown in the HSC-SSP Deep
images.

\section{Comparison of cluster properties in different catalogues}
\label{prevcatalog}

\begin{figure}
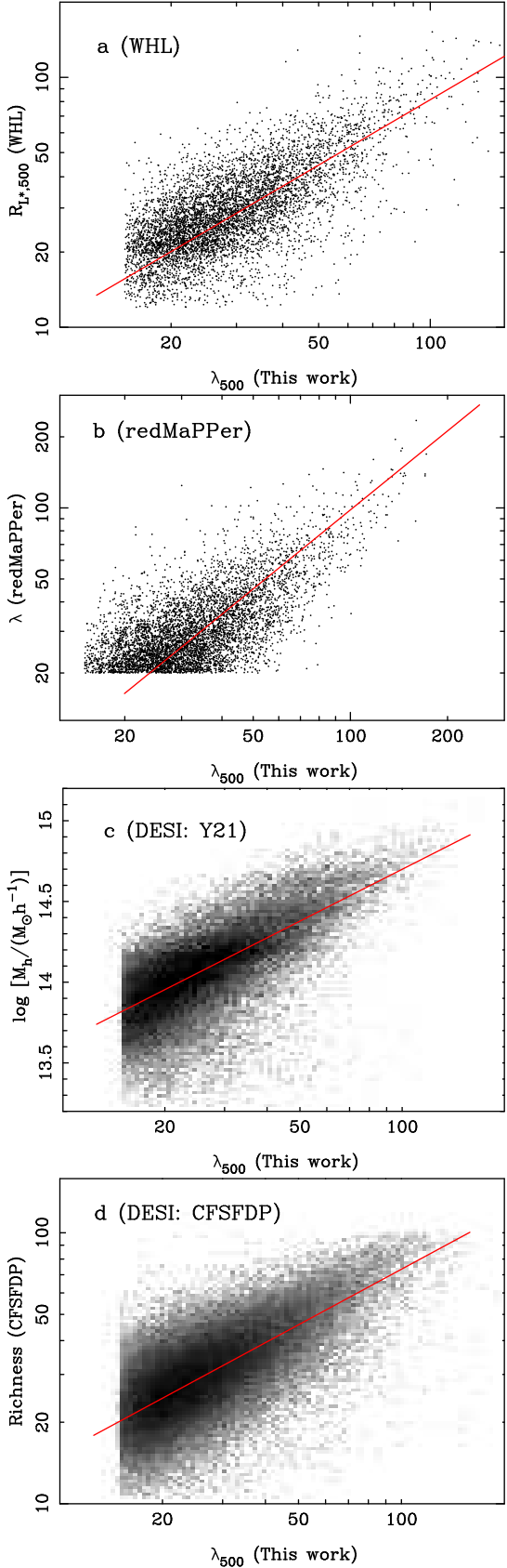

\centering \includegraphics[width = 0.4\textwidth]{f10a.eps}
\centering \includegraphics[width = 0.4\textwidth]{f10b.eps}
\centering \includegraphics[width = 0.4\textwidth]{f10c.eps}
\centering \includegraphics[width = 0.4\textwidth]{f10d.eps}
\caption{Comparison of our cluster richness with the richness (or halo
  mass) of WHL (panel a), redMaPPer (panel b), and DESI (panels c and d).
  The red lines are the best fits of the correlation.}
\label{richcomp}
\end{figure}

\begin{figure}
  \centering \includegraphics[width = 0.4\textwidth]{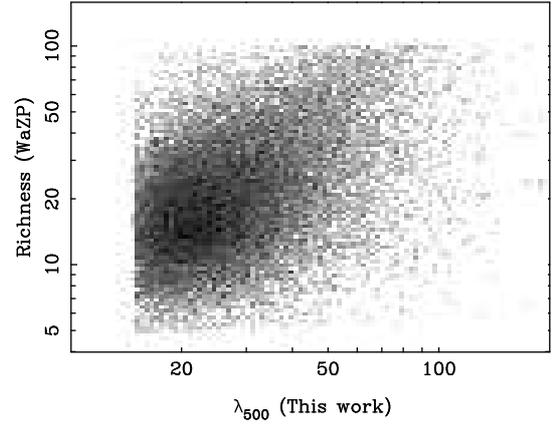}
\caption{Correlation of our cluster richness with the WaZP richness.}
\label{richwazp}
\end{figure}

Cluster finding algorithms have been applied to the DES data and
result in many catalogues of galaxy clusters, including the
red-sequence matched-filter Probabilistic Percolation cluster finder
\citep[redMaPPer;][]{rrb+14}, an extended halo-based group finder
\citep[Y21;][]{yxh+21}, clustering by Fast Search and Find of Density
Peaks \citep[CFSFDP;][]{zgx+21} and the Wavelet Z Photometric
\citep[WaZP;][]{abd+21}. In the DES footprint, a large sample of X-ray
clusters have been identified from the combined data of {\it ROSAT}
and the DES \citep{kgm+19}. Thousands of massive SZ clusters have been
identified up to a high redshift of $z>1$ from the {\it Planck}, SPT,
and ACT surveys \citep{planck16b,bbs+20,hsn+21}. Here, we compare the
properties of clusters in our catalogue from the DES with
those in previous catalogues.

\subsection{Comparison with optical clusters}

\subsubsection{WHL clusters}

Previously, we identified a sample of 158\,103 clusters from the
SDSS data using the similar algorithm \citep[][WHL
  hereafter]{whl12,wh15}. The richness is defined to be the total
luminosity of member galaxies within an estimated $r_{500}$.

There are 11685 WHL clusters of $z\ge0.1$ in the DES footprint, of
which 6360 (54 per cent) clusters have counterparts in this work within a redshift
difference of $0.05(1+z)$ and a projected distance of
$1.5\,r_{500}$. We find that the richnesses are well correlated in the two
catalogues (see Fig.~\ref{richcomp} a)
\begin{equation}
\log R_{\rm L*,500}=(0.87\pm0.01)\log \lambda_{500}+(0.17\pm0.02).
\end{equation}

\subsubsection{redMaPPer clusters}

The redMaPPer algorithm identifies clusters in multiband surveys
based on the red-sequence feature of cluster galaxies
\citep{rrb+14}. The algorithm first calibrates a red sequence model of
clusters as a function of redshift by using a training sample. Then,
the red sequence model is applied to photometric data to search
clusters. For each identified cluster, the algorithm estimates a membership
probability of red galaxies brighter than 0.2\,$L^{\ast}$ following an
iteratively matched-filter technique. The cluster richness is defined
to be the sum over the members with various probabilities. There are
6729 redMaPPer clusters at $0.2<z\lesssim 0.8$ in the DES Y1 data
covering a sky area of $\sim$1800 deg$^2$ \citep{rrh+16}.

Cross-matching shows that 5792 ($\sim$86 per cent) redMaPPer clusters are
found in Table~\ref{tab1} within a redshift difference of $0.05(1+z)$
and a projected distance of $1.5\,r_{500}$. The redMaPPer richness has
a good correlation with our richness (Fig.~\ref{richcomp} b), with the
best fit by a power law
\begin{equation}
\log \lambda_{500}=(0.90\pm0.01)\log \lambda ({\rm redMaPPer})+(0.21\pm0.02).
\end{equation}

\subsubsection{DESI clusters}

Dark Energy Spectroscopic Instrument (DESI) legacy imaging surveys are
composed of three optical surveys, i.e. the Beijing-Arizona Sky Survey
\citep{zzf+17}, the DES and the Mayall survey
\citep{zzz+18}. Recently, two large catalogues of clusters/groups have
been obtained from the data of DESI legacy imaging surveys
\citep{yxh+21,zgx+21}.

Using photometric or spectroscopic redshifts of galaxies,
\citet{yxh+21} applied an extend halo-based group finder to the DESI
legacy imaging surveys and identified 6.4 million groups with at least
three members. The halo masses of the groups are estimated based on
the total luminosities of group member galaxies. In the DES footprint,
152\,035 clusters of $z>0.1$ have at least 10 members. Our richness has
a tight correlation with their halo mass by a power law
(Fig.~\ref{richcomp} c),
\begin{equation}
\log M_h=(1.08\pm0.01)\log \lambda_{500}-(1.45\pm0.02),
\end{equation}
where $M_h$ is in units of $10^{14}~M_{\odot}h^{-1}$.

\citet{zgx+21} applied a similar algorithm to ours to have identified
galaxy clusters using photometric redshifts of galaxies in the DESI
legacy imaging surveys. They calculated the local density contrast of
galaxies within a photometric redshift slice to find the density peaks
as cluster candidates. The cluster richness is defined to be the total
luminosity of member galaxy candidates and is calibrated using a mass
known cluster sample. We find 56\,794 CFSFDP clusters detected in this
work and the CFSFDP richness is also well correlated with ours by a
power law (Fig.~\ref{richcomp} d),
\begin{equation}
\log \lambda(\rm CFSFDP)=(0.70\pm0.01)\log \lambda_{500}+(0.48\pm0.02).
\end{equation}

\subsubsection{WaZP clusters}

Using photometric redshifts of galaxies, \citet{abd+21} presented the
WaZP algorithm to identify clusters from DES Y1 data covering a sky
area of 1511.13 deg$^2$. The WaZP algorithm generates a smooth
wavelet-based density map within each photometric redshift slice. The
overdensity peaks are extracted from the smooth density maps as
cluster candidates. The cluster richness is defined to be the sum of
the membership probabilities for galaxies brighter than $m^{\ast}+1.5$
mag within a projected distance of $r_{\rm 200m}$. Here, $m^{\ast}$ is
the characteristic magnitude and $r_{\rm 200m}$ corresponds to the
radius which the mean density of a cluster is 200 times of the mean
density of the universe. The WaZP catalogue contains 60\,547
clusters at $0.05<z<0.9$, among which 20\,131 clusters are
detected in this work. However, the WaZP richness has a poor
correlation with our richness (Fig.~\ref{richwazp}). Some clusters
with a large value of WaZP richness are poor clusters in our catalogue.

\subsection{Comparison with X-ray clusters}

\begin{figure}
  \centering \includegraphics[width = 0.4\textwidth]{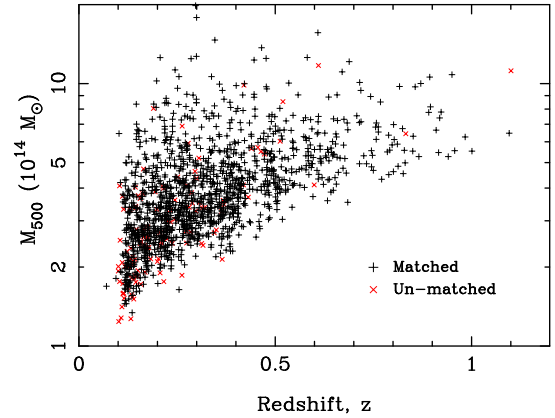}
  \caption{The matched and un-matched RASS X-ray clusters are shown
    in the redshift-mass diagram.}
\label{rass}
\end{figure}

Based on the second {\it ROSAT} All-Sky Survey (RASS) source catalogue
(2RXS) and the DES optical data, \citet{kgm+19} identified a sample
of X-ray clusters. Optical clusters are found by using a
multicomponent matched filter algorithm around the position of the
2RXS sources. A probability of being a random superposition, $f_{\rm
  cont}$, is estimated based on the redshift, optical richness and
implied X-ray luminosity. Their catalogue contains 1953 X-ray clusters
at redshifts $0.02<z<1.1$ if one sets a probability of $f_{\rm
  cont}<0.1$.
Among 1431 X-ray clusters of $z>0.1$, 1310 (92 per cent) clusters are matched
by clusters in our catalogue within a redshift difference of
$0.05(1+z)$ and a projected distance of $1.5\,r_{500}$. The
un-matched X-ray clusters in general have a lower mass (Fig.~\ref{rass}).

Our cluster catalogue can provide redshifts for many X-ray clusters.
The XMM Cluster Archive Super Survey (X-CLASS) is a serendipitous
search of X-ray clusters from {\it XMM--Newton} archival observations
\citep{csp+12}. The X-CLASS catalogue contains 1646 clusters up to
$z\sim1.5$, of which 281 clusters have no redshifts \citep{kcs+21}.
Cross-matching the 281 clusters with clusters in our catalogue, we get
the redshifts for 45 X-CLASS clusters, as listed in Table~\ref{taba1}.

\begin{figure}
  \centering \includegraphics[width = 0.4\textwidth]{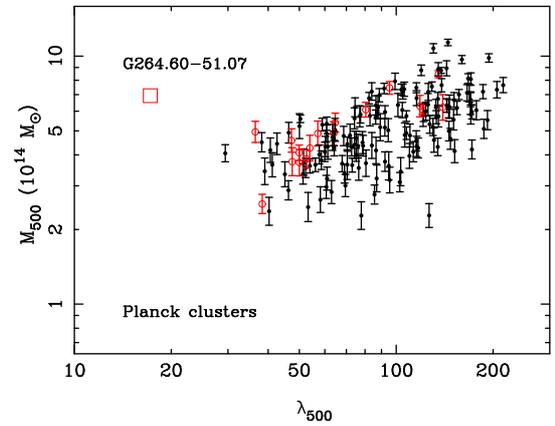}
  \caption{Correlation between cluster mass and our richness for
    {\it Planck} clusters. The circles indicate that the clusters have
    companions with a ratio of their richnesses less than 2. The 
    square indicates an outlier that a more massive foreground cluster
    is located along the line of sight of this {\it Planck} cluster.}
\label{matchplanck}
\end{figure}

\begin{figure}
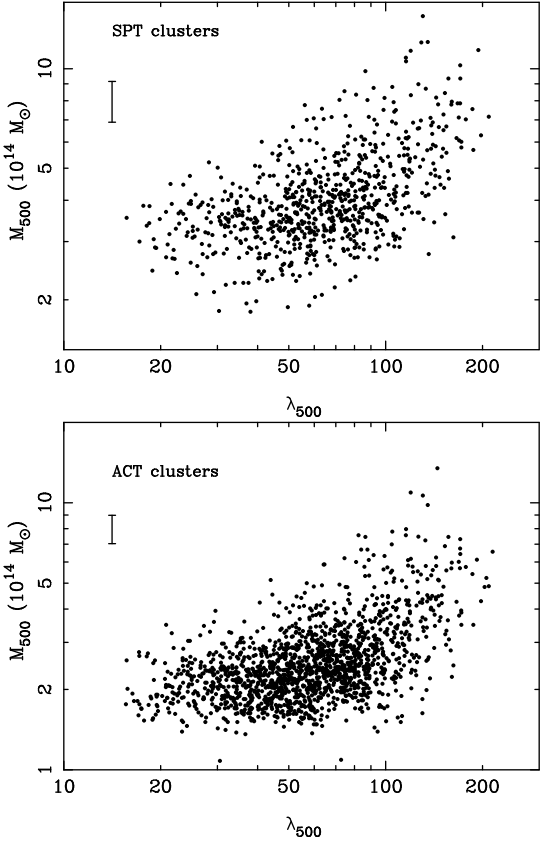

  \centering \includegraphics[width = 0.4\textwidth]{f14a.eps}
  \centering \includegraphics[width = 0.4\textwidth]{f14b.eps}
  \caption{Correlation between cluster mass and our richness for
    SPT (upper) and ACT (lower) clusters. The error bar indicates the
    typical uncertainty of the cluster mass in the SZ catalogues.}
\label{matchsptact}
\end{figure}

\subsection{Comparison with SZ clusters}

\subsubsection{{\it Planck} clusters}

The {\it Planck} satellite carries out an all sky millimeter survey
with an angular resolution of 5--10 arcmin in the millimeter bands
\citep{planck20}. The second release of the {\it Planck} SZ catalogue
contains 1653 clusters, most of which at redshifts $z<0.5$
\citep{planck16b}. There are 174 {\it Planck} clusters at $z>0.1$
in the DES footprint, of which 162 clusters are matched by clusters in our 
catalogue within a redshift difference of $0.05(1+z)$ and a projected
distance of $1.5\,r_{500}$. Among the remaining 12 un-matched
clusters, 10 clusters are found within a larger redshift difference
(see Appendix~\ref{plancknot}). If one also considers the above 10 clusters as
matches, our catalogue includes 99 per cent of the {\it Planck} clusters.

Fig.~\ref{matchplanck} shows a significant positive correlation
between our richness and the {\it Planck} cluster mass. Within of the
{\it Planck} beam, cluster pairs may be not resolved and both
contribute the SZ signal. We find that 23 clusters have a companion
cluster within a radius of 5 arcmin and with a ratio of their richness
less than 2. With contribution from companion clusters, the matched
clusters tend to have a larger mass estimate than other clusters with
the same richness. An outlier, PSZ2 G264.60$-$51.07, has a significant
offset from the correlation (red square). Careful checking shows that,
alone the line of sight towards this cluster, there is a more massive
foreground cluster, Abell 3128b at $z=0.06$ \citep{pap+11}.

\subsubsection{SPT clusters}
\label{matchspt}

The SPT observes the southern sky at the frequencies of 95, 150 and,
220 GHz with an angular resolution of 1--1.6 arcmin, making it is an
excellent facility to detect SZ clusters \citep{bsd+15}. Three surveys
have been carried out in different regions or with different depths,
including SPT-SZ, SPT Extended Cluster Survey (SPTECS), and SPTpol
100d. The SPT-SZ and SPTECS cluster catalogues contain 677 clusters
from a sky area of 2500 deg$^2$, 483 clusters from a sky area of 2770
deg$^2$, respectively \citep{bsd+15,bbs+20}. The SPTpol 100d cluster
catalogue contains 89 clusters extending to a lower mass than the
SPT-SZ and SPT-ECS \citep{hbs+20}.

In the DES footprint, we get 844 unique SPT clusters at redshifts
$0.1<z\lesssim 1.5$, of which 720 clusters are matched by clusters in our
catalogue within a redshift difference of $0.05(1+z)$ and a projected
distance of $1.5\,r_{500}$. Another 78 SPT clusters can be matched
within a redshift difference of $0.05(1+z)$--$0.1(1+z)$ from our
clusters. Among the remaining 46 un-matched SPT clusters, four of
them can be matched within a larger projected distance of
$1.5\,r_{500}$--$2\,r_{500}$. Visual inspection shows that these
clusters are likely in double cluster systems. Another 18 clusters can
be matched within a larger redshift difference. If one considers these
clusters as matches, our catalogue includes 97 per cent of the SPT
clusters. Fig.~\ref{matchsptact} (upper panel) shows that our richness
has a positive correlation with the SPT cluster mass.

Redshifts of 174 SPT clusters in the DES footprint are unknown and
they can be determined by using data in our catalogue. For an SZ
cluster located within a projected distance of $1.5\,r_{500}$ from an
optical cluster, we take the redshift of the optical cluster. Through
this procedure, we get redshifts for 56 SPT clusters, as listed in
Table~\ref{taba2}, of which 15 clusters have redshifts $z>1$.

\subsubsection{ACT clusters}

The ACT is designed to detect galaxy clusters at the frequencies of
148 GHz, 218 GHz, and 227 GHz with an angular resolution of 1.4 arcmin
at 148 GHz \citep{saa+11}. The latest ACT SZ catalogue includes 4195
clusters at redshifts $0.04<z<1.95$ from a sky area of 13,211 deg$^2$
\citep{hsn+21}, of which 1819 clusters are located in the DES
footprint. We find that 1698 (93 per cent) ACT clusters are matched by
clusters in our catalogue within a redshift difference of
$0.1(1+z)$ and a projected distance of $1.5\,r_{500}$. The un-matched
matched ACT clusters generally have a low mass. Careful checking the
un-matched clusters, we find another eight clusters can be matched by
our optical clusters in double cluster systems and 17 clusters can be
matched within a redshift difference larger than $0.1(1+z)$. Therefore
95 per cent of ACT clusters can be matched by our clusters. A positive
correlation is also found between our richness and the ACT cluster
mass (the lower panel of Fig.~\ref{matchsptact}).

\subsubsection{Multiple clusters along a line of sight}

\begin{figure}
  \centering \includegraphics[width = 0.4\textwidth]{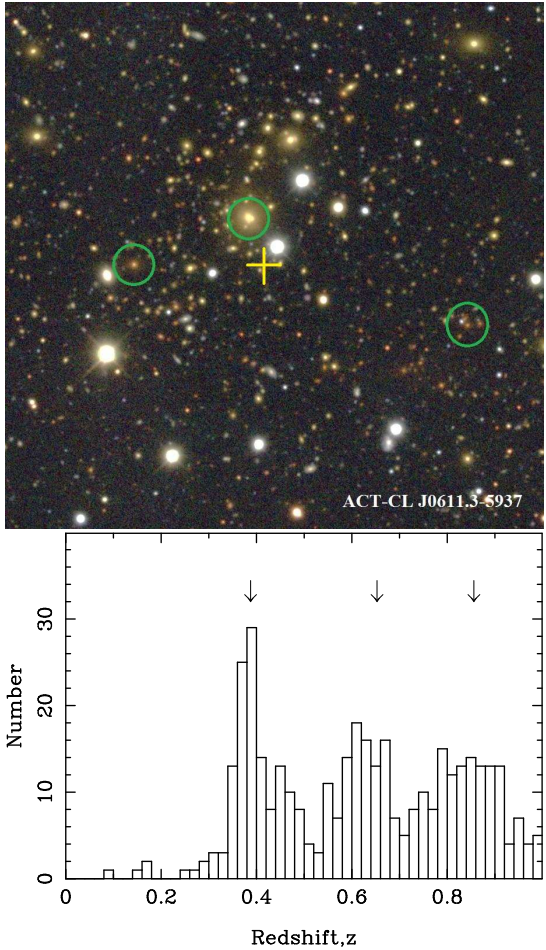}
  \centering \includegraphics[width = 0.4\textwidth]{f15b.eps}
  \caption{Optical image for an example of multiple clusters along the line of sight
    towards an SZ cluster, ACT-CL J0611.3$-$5937, at $z=0.39$.
    The image has a size of 4 arcmin $\times$ 4 arcmin. In the
    upper panel, green circles indicate three optical clusters
    (centring on BCGs) we identify at $z=0.3643$, 0.6295 and 0.8423,
    respectively. The yellow plus indicates the position of the SZ
    cluster. The lower panel
    shows the photometric redshift distribution of galaxies within a
    projected radius 1.5 Mpc from the SZ position. The arrows
    indicate the redshifts of three optical clusters.}
\label{overlap}
\end{figure}

Multiple clusters can be found along a line of sight due to the
projection effect, which may induce bias on cluster mass by SZ and
weak lensing effects \citep{whs02,shb08,ylp+18}. Here, we investigate
how many SZ massive clusters match multiple optical clusters in the
line of sight. An optical cluster is regarded as the counterpart in
the line of sight towards an SZ cluster if it is located within a
projected distance of $r_{500}$. We find multiple optical clusters in
the line of sight towards 155 SPT clusters and 236 ACT clusters,
respectively, as listed in Table~\ref{taba3}. Therefore, multiple
optical clusters can be found in the line of sight towards about 15 per cent
of SZ clusters.

Fig.~\ref{overlap} shows an example that multiple
optical clusters located along the line of sight towards an SZ
cluster, ACT-CL J0611.3$-$5937, at $z=0.39$. Three peaks are clearly
shown in the photometric redshift distribution of galaxies,
corresponding three clusters at redshifts of 0.3643, 0.6295, and
0.8423 with richnesses of 76.00, 24.88, and 59.17,
respectively.

\section{Summary}

We present photometric redshifts for galaxies in the DES$\times$unWISE
data and identify a large sample of galaxy clusters. Using the
nearest-neighbour algorithm, we first estimate the photometric
redshifts for 105 million DES$\times$unWISE galaxies. The
redshift uncertainty increases from 0.013 at $z\sim0.1$ to 0.052 at
$z\sim1.3$. The redshift distribution has a peak at $z\sim 0.7$ and
extends to $z\sim 1.7$. The stellar mass is derived from the
mid-infrared luminosity.
Then, we identify a large sample of 151\,244 clusters at redshifts
$0.1<z\lesssim 1.5$ based on the overdensity of galaxy stellar mass
within a photometric redshift slice. The cluster redshift uncertainty
is about 0.013 at redshifts $z<0.9$. Compared to the published DES
cluster catalogues, our catalogue extends to a much higher
redshift. We cross-match our clusters with SZ clusters in the {\it
  Planck}, SPT and ACT surveys, and X-ray clusters. More than 95 percent SZ
clusters can be matched with clusters in our catalogue. Redshifts are
well determined for 45 X-ray clusters and 56 SZ clusters. We find
multiple optical clusters in the line of sight towards about 15 percent of
SZ clusters.

\section*{Acknowledgements}

We thank the referee for valuable comments that helped to improve the
paper. The authors are partially supported by the National Natural
Science Foundation of China (Grant Numbers 11988101, 11833009, 12073036 and
U1931202), the Key Research Program of the Chinese Academy of
Sciences (Grant Number QYZDJ-SSW-SLH021). We also acknowledge the
support by the science research grants from the China Manned Space
Project with Numbers CMS-CSST-2021-A01, CMS-CSST-2021-A04 and
CMS-CSST-2021-B01.

This project used public archival data from the Dark Energy Survey
(DES). Funding for the DES Projects has been provided by the
U.S. Department of Energy, the U.S. National Science Foundation, the
Ministry of Science and Education of Spain, the Science and Technology
FacilitiesCouncil of the United Kingdom, the Higher Education Funding
Council for England, the National Center for Supercomputing
Applications at the University of Illinois at Urbana-Champaign, the
Kavli Institute of Cosmological Physics at the University of Chicago,
the Center for Cosmology and Astro-Particle Physics at the Ohio State
University, the Mitchell Institute for Fundamental Physics and
Astronomy at Texas A\&M University, Financiadora de Estudos e
Projetos, Funda{\c c}{\~a}o Carlos Chagas Filho de Amparo {\`a}
Pesquisa do Estado do Rio de Janeiro, Conselho Nacional de
Desenvolvimento Cient{\'i}fico e Tecnol{\'o}gico and the
Minist{\'e}rio da Ci{\^e}ncia, Tecnologia e Inova{\c c}{\~a}o, the
Deutsche Forschungsgemeinschaft, and the Collaborating Institutions in
the Dark Energy Survey.
The Collaborating Institutions are Argonne National Laboratory, the
University of California at Santa Cruz, the University of Cambridge,
Centro de Investigaciones Energ{\'e}ticas, Medioambientales y
Tecnol{\'o}gicas-Madrid, the University of Chicago, University College
London, the DES-Brazil Consortium, the University of Edinburgh, the
Eidgen{\"o}ssische Technische Hochschule (ETH) Z{\"u}rich, Fermi
National Accelerator Laboratory, the University of Illinois at
Urbana-Champaign, the Institut de Ci{\`e}ncies de l'Espai (IEEC/CSIC),
the Institut de F{\'i}sica d'Altes Energies, Lawrence Berkeley
National Laboratory, the Ludwig-Maximilians Universit{\"a}t
M{\"u}nchen and the associated Excellence Cluster Universe, the
University of Michigan, the National Optical Astronomy Observatory,
the University of Nottingham, The Ohio State University, the OzDES
Membership Consortium, the University of Pennsylvania, the University
of Portsmouth, SLAC National Accelerator Laboratory, Stanford
University, the University of Sussex, and Texas A\&M University.
This publication makes use of data products from the Wide-field
Infrared Survey Explorer, which is a joint project of the University
of California, Los Angeles, and the Jet Propulsion
Laboratory/California Institute of Technology, funded by the National
Aeronautics and Space Administration.
This work makes use of data based on observations made with ESO
Telescopes at the La Silla or Paranal Observatories under programme
ID(s) 179.A-2004(M) and 179.A-2010(A).

\section*{Data availability}

The data of 105 million DES$\times$unWISE galaxies (seven-band
magnitudes, photometric redshifts, and stellar masses), 151\,244
identified clusters and their 2\,454\,302 member galaxy candidates are
publicly available at http://zmtt.bao.ac.cn/galaxy\_clusters/.

\bibliographystyle{mnras}
\bibliography{des}

\begin{appendix}

\section{Tables of cross-matching with X-ray and SZ cluster catalogues}

In Section 4.2--4.3, we cross-match the clusters in our catalogue with
clusters in X-ray and SZ cluster catalogues. We determine redshifts
for 45 X-CLASS clusters and 56 SPT clusters, as listed in
Tables~\ref{taba1} and \ref{taba2}, respectively. In
Table~\ref{taba3}, we list multiple optical clusters in the line of
sight towards SZ clusters.

\begin{table*}
\begin{minipage}{160mm}
\caption[]{45 X-CLASS clustes with redshifts determined based on our cluster catalogue.}
\begin{center}
\setlength{\tabcolsep}{1mm}
\begin{tabular}{rrrrrrcrc}
\hline
\mc{1}{c}{X-CLASS ID}& \mc{1}{c}{R.A.$_{\rm X}$} & \mc{1}{c}{Dec.$_{\rm X}$} & \mc{1}{c}{MLdet} &
\mc{1}{c}{R.A.} & \mc{1}{c}{Dec.} &  \mc{1}{c}{$z_{\rm cl}$} & \mc{1}{c}{$\lambda_{500}$}   &\mc{1}{c}{$r_p$} \\
\mc{1}{c}{(1)} & \mc{1}{c}{(2)} & \mc{1}{c}{(3)} & \mc{1}{c}{(4)} & \mc{1}{c}{(5)} & 
\mc{1}{c}{(6)} & \mc{1}{c}{(7)} & \mc{1}{c}{(8)} & \mc{1}{c}{(9)} \\
\hline
   551&   5.6187& $-48.7262$ &   64.52 &   5.62022 & $-48.72610$ & 0.9226 & 33.72 & 0.028 \\
  2716&   6.3299& $-26.1939$ &  114.75 &   6.31623 & $-26.24651$ & 0.4640 & 68.61 & 1.140 \\
  2226&  14.2757& $-26.3405$ &   62.70 &  14.27520 & $-26.33146$ & 0.4165 & 44.78 & 0.179 \\
  3138&  15.7980& $-22.2136$ &  141.56 &  15.82439 & $-22.19425$ & 0.2221 & 26.57 & 0.402 \\
   222&  16.0734& $ -6.7455$ &   75.44 &  16.07304 & $ -6.74599$ & 0.5827 & 22.86 & 0.014 \\
   132&  19.9576& $-44.0768$ &  263.79 &  19.95924 & $-44.07513$ & 0.3732 & 25.67 & 0.038 \\
    98&  23.0731& $-13.7561$ &  112.05 &  23.07206 & $-13.74955$ & 0.5694 & 42.60 & 0.156 \\
  3271&  23.6633& $-29.5465$ &  101.07 &  23.64802 & $-29.51678$ & 0.4248 & 25.40 & 0.653 \\
  2958&  24.3381& $ -8.5275$ &   81.63 &  24.33810 & $ -8.52086$ & 0.7872 & 35.92 & 0.178 \\
 22690&  29.9973& $-40.5389$ &  856.32 &  29.99760 & $-40.53904$ & 0.7631 & 37.79 & 0.007 \\
    59&  31.9576& $  2.1567$ &  302.38 &  31.95749 & $  2.15527$ & 0.8374 & 19.82 & 0.039 \\
 22724&  38.0413& $-57.7660$ &  416.02 &  38.04234 & $-57.76540$ & 0.5591 & 39.13 & 0.019 \\
  2831&  38.3807& $-58.3136$ &   54.14 &  38.37849 & $-58.31344$ & 0.6689 & 36.73 & 0.030 \\
  1998&  40.1097& $-34.5564$ &  164.90 &  40.10824 & $-34.55753$ & 0.2459 & 24.92 & 0.023 \\
  3357&  41.4938& $-20.4883$ & 4011.19 &  41.49205 & $-20.48772$ & 0.3476 & 62.24 & 0.031 \\
  2475&  52.7173& $-38.2198$ &  229.83 &  52.71795 & $-38.21850$ & 0.3526 & 39.19 & 0.025 \\
  2476&  52.8683& $-38.4078$ &  262.73 &  52.88900 & $-38.39761$ & 0.5693 & 41.33 & 0.450 \\
  2477&  52.9013& $-38.5544$ &   71.18 &  52.90094 & $-38.55338$ & 0.6817 & 73.57 & 0.027 \\
   380&  54.8022& $-35.4366$ &  586.46 &  54.82275 & $-35.41658$ & 0.5213 & 24.76 & 0.586 \\
  3246&  55.0276& $-28.8449$ & 1174.80 &  55.03021 & $-28.84417$ & 0.3412 & 51.92 & 0.042 \\
  3396&  55.2673& $-18.4472$ &  121.58 &  55.29052 & $-18.45590$ & 0.6856 & 39.80 & 0.604 \\
 24804&  55.3873& $-54.0582$ &  167.08 &  55.38786 & $-54.05875$ & 0.5166 & 30.51 & 0.014 \\
 24801&  55.9383& $-53.2460$ &  354.53 &  55.93827 & $-53.24508$ & 0.6967 & 48.00 & 0.024 \\
  3382&  56.0465& $-54.8537$ &  139.06 &  56.04767 & $-54.85653$ & 1.0766 & 34.74 & 0.085 \\
  2833&  56.2025& $-54.8733$ &   48.01 &  56.20447 & $-54.87134$ & 0.3837 & 28.47 & 0.043 \\
 22696&  58.8494& $-59.0921$ &  365.78 &  58.79378 & $-59.09582$ & 0.7336 & 38.78 & 0.754 \\
 21748&  61.9848& $-21.0035$ &   85.82 &  61.98438 & $-21.00267$ & 0.2878 & 23.65 & 0.014 \\
  3469&  63.7497& $-58.7290$ &  276.26 &  63.71315 & $-58.73209$ & 0.4949 & 28.27 & 0.420 \\
   461&  76.2676& $-28.7829$ &   58.70 &  76.26675 & $-28.78091$ & 0.5255 & 22.25 & 0.048 \\
   395&  80.8841& $-36.6406$ &  108.70 &  80.88306 & $-36.64333$ & 1.0064 & 53.96 & 0.082 \\
  2663&  85.8359& $-35.9356$ &  474.65 &  85.83328 & $-35.93603$ & 0.5929 & 18.24 & 0.052 \\
  3466&  89.3000& $-33.0104$ &  201.86 &  89.27305 & $-32.96622$ & 0.1443 & 39.35 & 0.452 \\
 22420&  94.0889& $-39.6170$ &  579.30 &  94.08949 & $-39.61694$ & 0.1560 & 22.14 & 0.004 \\
 22740& 302.7220& $-57.3814$ &  799.61 & 302.72629 & $-57.38272$ & 0.6469 & 89.72 & 0.066 \\
  3366& 304.6980& $-62.9852$ &  161.28 & 304.69986 & $-62.98596$ & 0.3655 & 18.49 & 0.021 \\
  3363& 308.1060& $-56.3725$ &   65.94 & 308.05853 & $-56.43681$ & 0.2955 & 82.10 & 1.102 \\
 24648& 309.6170& $-56.2482$ &   81.84 & 309.60733 & $-56.25622$ & 0.3723 & 64.57 & 0.179 \\
 24502& 314.7920& $-42.8471$ &  121.67 & 314.79187 & $-42.84774$ & 0.3465 & 56.33 & 0.011 \\
 24503& 314.9620& $-42.8804$ &   88.47 & 314.96161 & $-42.87986$ & 1.1197 & 28.00 & 0.018 \\
 22709& 322.6650& $ -0.2846$ &   34.33 & 322.66156 & $ -0.28537$ & 0.7476 & 27.48 & 0.093 \\
  3376& 324.0400& $-63.1761$ &  250.59 & 324.03592 & $-63.17659$ & 0.2299 & 46.36 & 0.025 \\
  2814& 326.5680& $-56.7574$ &  109.14 & 326.59476 & $-56.77512$ & 0.4954 & 44.40 & 0.503 \\
  2815& 326.6770& $-56.8749$ &  237.72 & 326.67700 & $-56.87627$ & 0.4901 & 42.57 & 0.030 \\
 24791& 355.1730& $-51.4972$ &   73.98 & 355.17575 & $-51.50106$ & 0.9692 & 20.81 & 0.121 \\
  2651& 355.3030& $-60.9058$ &  158.14 & 355.33646 & $-60.90015$ & 0.5047 & 21.16 & 0.380 \\ 
 \hline
\end{tabular}
\end{center}
{Note.
Column 1: ID in X-CLASS catalogue; 
Column 2 and 3: Right Ascension (R.A. J2000) and Declination (Dec. J2000) of cluster (in degree);
Column 4: X-ray detection likelihood; 
Column 5--6: Right Ascension (R.A. J2000) and Declination (Dec. J2000) of the cluster in our catalogue;
Column 7: redshift;
Column 8: richness.
Column 9: projected separation between the cluster centres in two catalogues, in Mpc.\\
}

\label{taba1}
\end{minipage}
\end{table*}

\begin{table*}
\begin{minipage}{160mm}
\caption[]{56 SPT clustes with redshifts determined.}
\begin{center}
\setlength{\tabcolsep}{1mm}
\begin{tabular}{crrcrrcrc}
\hline
\mc{1}{c}{SZ cluster name}& \mc{1}{c}{R.A.$_{\rm SZ}$} & \mc{1}{c}{Dec.$_{\rm SZ}$} & \mc{1}{c}{$\xi$} &
\mc{1}{c}{R.A.} & \mc{1}{c}{Dec.} &  \mc{1}{c}{$z_{\rm cl}$} & \mc{1}{c}{$\lambda_{500}$}   &\mc{1}{c}{$r_p$} \\
\mc{1}{c}{(1)} & \mc{1}{c}{(2)} & \mc{1}{c}{(3)} & \mc{1}{c}{(4)} & \mc{1}{c}{(5)} & 
\mc{1}{c}{(6)} & \mc{1}{c}{(7)} & \mc{1}{c}{(8)} & \mc{1}{c}{(9)} \\
\hline
 SPT-CLJ0013-5714 &      3.3029& $-57.2373$ &  5.11 &   3.27961 & $-57.23208$ & 0.6061 & 27.16& 0.330 \\
 SPT-CLJ0013-5310 &      3.4111& $-53.1718$ &  4.66 &   3.40160 & $-53.17379$ & 0.9003 & 54.96& 0.169 \\
 SPT-CLJ0030-4457 &      7.5150& $-44.9640$ &  4.68 &   7.53638 & $-44.98155$ & 0.5585 & 27.02& 0.539 \\
 SPT-CLJ0034-5554 &      8.5906& $-55.9013$ &  4.54 &   8.57116 & $-55.89748$ & 0.9601 & 39.62& 0.329 \\
 SPT-CLJ0048-4450 &     12.1742& $-44.8475$ &  4.82 &  12.17938 & $-44.84469$ & 0.5950 & 49.17& 0.111 \\
 SPT-CLJ0101-3109 &     15.2898& $-31.1514$ &  5.19 &  15.30829 & $-31.13954$ & 0.5543 & 17.85& 0.458 \\
 SPT-CLJ0105-5358 &     16.3908& $-53.9717$ &  4.53 &  16.36311 & $-53.97701$ & 0.7103 & 20.78& 0.443 \\
 SPT-CLJ0112-5030 &     18.0354& $-50.5074$ &  4.50 &  18.00619 & $-50.50797$ & 0.6775 & 19.25& 0.472 \\
 SPT-CLJ0144-4157 &     26.1463& $-41.9601$ &  5.16 &  26.14669 & $-41.95748$ & 1.0472 & 69.93& 0.077 \\
 SPT-CLJ0146-6126 &     26.6503& $-61.4365$ &  4.87 &  26.68257 & $-61.42861$ & 0.8602 & 46.77& 0.479 \\
 SPT-CLJ0151-4300 &     27.8277& $-43.0004$ &  4.56 &  27.82843 & $-43.00207$ & 1.1521 & 76.42& 0.052 \\
 SPT-CLJ0220-2825 &     35.0214& $-28.4171$ &  5.14 &  35.03450 & $-28.42431$ & 1.1401 & 20.47& 0.402 \\
 SPT-CLJ0220-5445 &     35.1200& $-54.7511$ &  4.57 &  35.13049 & $-54.75689$ & 0.7007 & 34.10& 0.216 \\
 SPT-CLJ0244-4755 &     41.0159& $-47.9274$ &  4.79 &  41.00558 & $-47.89097$ & 0.2604 & 28.77& 0.538 \\
 SPT-CLJ0250-4714 &     42.6656& $-47.2385$ &  4.50 &  42.66674 & $-47.25129$ & 0.4737 & 32.74& 0.273 \\
 SPT-CLJ0256-4221 &     44.0827& $-42.3589$ &  4.54 &  44.06901 & $-42.36617$ & 0.8458 & 16.90& 0.343 \\
 SPT-CLJ0323-4913 &     50.9166& $-49.2215$ &  4.54 &  50.89778 & $-49.19088$ & 0.3881 & 19.01& 0.627 \\
 SPT-CLJ0330-2016 &     52.6254& $-20.2713$ &  5.29 &  52.62220 & $-20.25523$ & 1.2460 & 24.27& 0.490 \\
 SPT-CLJ0337-4442 &     54.2698& $-44.7105$ &  4.84 &  54.26978 & $-44.70407$ & 0.9508 & 59.78& 0.183 \\
 SPT-CLJ0337-6207 &     54.4708& $-62.1175$ &  5.03 &  54.47217 & $-62.10662$ & 0.9633 & 53.04& 0.311 \\
 SPT-CLJ0339-3952 &     54.7862& $-39.8734$ &  5.45 &  54.78075 & $-39.86796$ & 1.2059 & 65.61& 0.205 \\
 SPT-CLJ0353-4818 &     58.2639& $-48.3160$ &  4.90 &  58.25852 & $-48.31760$ & 1.0964 & 32.45& 0.115 \\
 SPT-CLJ0353-5312 &     58.3058& $-53.2095$ &  4.54 &  58.32022 & $-53.19808$ & 0.6205 & 58.81& 0.350 \\
 SPT-CLJ0353-5043 &     58.3853& $-50.7278$ &  5.35 &  58.37934 & $-50.73331$ & 1.2203 & 23.84& 0.200 \\
 SPT-CLJ0430-6251 &     67.7094& $-62.8548$ &  5.29 &  67.75693 & $-62.86843$ & 0.5373 & 27.84& 0.584 \\
 SPT-CLJ0437-5307 &     69.2599& $-53.1206$ &  4.52 &  69.26513 & $-53.09996$ & 0.2885 & 37.92& 0.326 \\
 SPT-CLJ0442-4309 &     70.6940& $-43.1528$ &  4.63 &  70.74709 & $-43.13006$ & 0.4334 & 21.00& 0.912 \\
 SPT-CLJ0447-5041 &     71.7821& $-50.6991$ &  4.62 &  71.79196 & $-50.67995$ & 1.3027 & 19.75& 0.607 \\
 SPT-CLJ0448-4036 &     72.0160& $-40.6031$ &  4.66 &  72.02155 & $-40.60420$ & 1.0942 & 53.79& 0.128 \\
 SPT-CLJ0455-2225 &     73.9549& $-22.4246$ &  5.04 &  73.95729 & $-22.45580$ & 0.4244 & 28.21& 0.627 \\
 SPT-CLJ0500-4713 &     75.1198& $-47.2293$ &  4.92 &  75.09506 & $-47.23772$ & 1.2506 & 35.88& 0.563 \\
 SPT-CLJ0501-4455 &     75.2932& $-44.9270$ &  4.59 &  75.26968 & $-44.93023$ & 0.4557 & 83.97& 0.354 \\
 SPT-CLJ0502-6048 &     75.7229& $-60.8112$ &  4.69 &  75.69482 & $-60.79540$ & 0.8068 & 18.28& 0.567 \\
 SPT-CLJ0534-4847 &     83.6655& $-48.7965$ &  4.59 &  83.67176 & $-48.79736$ & 0.7860 & 58.92& 0.113 \\
 SPT-CLJ0545-5052 &     86.4957& $-50.8814$ &  4.60 &  86.49717 & $-50.87030$ & 0.9295 & 25.43& 0.315 \\
 SPT-CLJ0546-6040 &     86.7342& $-60.6723$ &  4.72 &  86.73521 & $-60.68235$ & 0.8388 & 27.16& 0.276 \\
 SPT-CLJ0548-4340 &     87.2419& $-43.6819$ &  4.59 &  87.24103 & $-43.68504$ & 0.8307 & 24.51& 0.088 \\
 SPT-CLJ0552-4937 &     88.1783& $-49.6208$ &  5.05 &  88.18316 & $-49.60241$ & 0.9929 & 23.40& 0.537 \\
 SPT-CLJ0559-6022 &     89.9419& $-60.3832$ &  5.67 &  89.94439 & $-60.38390$ & 1.1109 & 38.83& 0.042 \\
 SPT-CLJ0601-5204 &     90.2515& $-52.0797$ &  4.54 &  90.25552 & $-52.08411$ & 1.0072 & 24.61& 0.146 \\
 SPT-CLJ0611-6000 &     92.7794& $-60.0087$ &  4.68 &  92.80997 & $-60.01533$ & 0.6466 & 25.56& 0.414 \\
 SPT-CLJ2035-5614 &    308.9018& $-56.2402$ &  4.62 & 308.89615 & $-56.23104$ & 0.7964 & 47.34& 0.261 \\
 SPT-CLJ2106-5820 &    316.5130& $-58.3467$ &  4.65 & 316.60272 & $-58.37299$ & 0.2621 & 36.30& 0.786 \\
 SPT-CLJ2121-5546 &    320.2690& $-55.7782$ &  4.57 & 320.27945 & $-55.78068$ & 1.0916 & 87.17& 0.187 \\
 SPT-CLJ2136-5723 &    324.1203& $-57.3968$ &  4.72 & 324.17242 & $-57.41769$ & 0.3057 & 32.52& 0.569 \\
 SPT-CLJ2139-6430 &    324.9603& $-64.5095$ &  4.53 & 324.95728 & $-64.51631$ & 0.9881 & 58.74& 0.199 \\
 SPT-CLJ2152-4629 &    328.1926& $-46.4954$ &  4.83 & 328.18686 & $-46.49863$ & 0.9300 & 34.65& 0.144 \\
 SPT-CLJ2158-4851 &    329.5692& $-48.8533$ &  4.64 & 329.55301 & $-48.84221$ & 0.5149 & 35.83& 0.343 \\
 SPT-CLJ2233-4729 &    338.3648& $-47.4912$ &  4.59 & 338.33438 & $-47.49746$ & 1.1691 & 26.09& 0.638 \\
 SPT-CLJ2300-5148 &    345.0616& $-51.8003$ &  4.92 & 345.03333 & $-51.78909$ & 0.8232 & 49.98& 0.567 \\
 SPT-CLJ2310-5239 &    347.7022& $-52.6609$ &  4.91 & 347.70789 & $-52.67169$ & 0.3628 & 15.49& 0.206 \\
 SPT-CLJ2328-4616 &    352.0576& $-46.2802$ &  4.84 & 351.98587 & $-46.25559$ & 0.2117 & 28.68& 0.688 \\
 SPT-CLJ2339-5008 &    354.9618& $-50.1427$ &  4.81 & 354.95956 & $-50.14320$ & 0.9275 & 21.57& 0.043 \\
 SPT-CLJ2352-5846 &    358.0510& $-58.7758$ &  5.18 & 358.15778 & $-58.74107$ & 0.1692 & 24.63& 0.680 \\
 SPT-CLJ2357-5953 &    359.2865& $-59.8988$ &  4.67 & 359.31033 & $-59.90747$ & 0.5114 & 22.40& 0.328 \\
 SPT-CLJ2358-4143 &    359.6421& $-41.7173$ &  4.74 & 359.61539 & $-41.73634$ & 1.1105 & 36.12& 0.812 \\ 
 \hline
\end{tabular}
\end{center}
{Note.
Column 1: Name of SPT cluster; 
Column 2 and 3: Right Ascension (R.A. J2000) and Declination (Dec. J2000) of SPT cluster (in degree);
Column 4: SZ detection significane; 
Column 5--6: Right Ascension (R.A. J2000) and Declination (Dec. J2000) of the cluster in our catalogue;
Column 7: redshift;
Column 8: richness.
Column 9: projected separation between the cluster centres in two catalogues, in Mpc.\\
}
\label{taba2}
\end{minipage}
\end{table*}

\begin{table*}
\begin{minipage}{160mm}
\caption[]{Examples for multiple clusters in the line of sight towards 155 SPT clusters and 236 ACT clusters.}
\begin{center}
\setlength{\tabcolsep}{1mm}
\begin{tabular}{crrcrrcrcc}
\hline
\mc{1}{c}{SZ cluster name}& \mc{1}{c}{R.A.$_{\rm SZ}$} & \mc{1}{c}{Dec.$_{\rm SZ}$} & \mc{1}{c}{SNR} & \mc{1}{c}{$z_{\rm SZ}$} &
\mc{1}{c}{R.A.} & \mc{1}{c}{Dec.} &  \mc{1}{c}{$z_c$} & \mc{1}{c}{$\lambda_{500}$}   &\mc{1}{c}{$\theta$} \\
\mc{1}{c}{(1)} & \mc{1}{c}{(2)} & \mc{1}{c}{(3)} & \mc{1}{c}{(4)} & \mc{1}{c}{(5)} & 
\mc{1}{c}{(6)} & \mc{1}{c}{(7)} & \mc{1}{c}{(8)} & \mc{1}{c}{(9)} & \mc{1}{c}{(10)} \\
\hline
 SPT-CLJ0010-5112   &    2.7408& $-51.2077$ &  4.51 & 0.1700 &   2.74925 & $-51.20449$ & 0.2077 & 47.05 &  0.37\\
                    &          &            &       &        &   2.86313 & $-51.20541$ & 0.7615 & 18.46 &  4.60\\
 SPT-CLJ0012-5352   &    3.0649& $-53.8736$ &  4.80 & 0.3300 &   3.06954 & $-53.85092$ & 0.3488 & 66.51 &  1.37\\
                    &          &            &       &        &   3.08529 & $-53.85501$ & 0.8924 & 35.06 &  1.33\\
 SPT-CLJ0013-4621   &    3.4715& $-46.3563$ &  4.54 & 0.1800 &   3.53073 & $-46.31323$ & 0.1996 & 37.06 &  3.56\\
                    &          &            &       &        &   3.44134 & $-46.32892$ & 0.9663 & 48.29 &  2.06\\
                    &          &            &       &$\cdots\cdots$&     &             &        &       &      \\
 ACT-CLJ0000.7+0225 &    0.1791& $  2.4184$ &  6.19 & 0.4227 &   0.17833 & $  2.42094$ & 0.4268 & 29.59 &  0.16\\
                    &          &            &       &        &   0.15072 & $  2.40283$ & 1.0894 & 18.10 &  1.94\\
 ACT-CLJ0012.3-5352 &    3.0875& $-53.8750$ &  4.14 & 0.3431 &   3.06954 & $-53.85092$ & 0.3488 & 66.51 &  1.58\\
                    &          &            &       &        &   3.08529 & $-53.85501$ & 0.8924 & 35.06 &  1.20\\
 ACT-CLJ0014.2-3348 &    3.5504& $-33.8014$ &  6.37 & 0.4007 &   3.52542 & $-33.81221$ & 0.3922 & 38.11 &  1.40\\
                    &          &            &       &        &   3.55809 & $-33.80380$ & 0.5370 & 15.68 &  0.41\\ 
 \hline
\end{tabular}
\end{center}
{Note.
Column 1: Name of SZ cluster;
Column 2 and 3: Right Ascension (R.A. J2000) and Declination (Dec. J2000) of SZ cluster (in degree);
Column 4: the significance of the SZ detection;
Column 5: redshift of SZ cluster; 
Column 6--7: Right Ascension and Declination of a cluster in our catalogue;
Column 8: redshift of the optical cluster;
Column 9: cluster richness estimated from the stellar mass.
Column 10: angular distance between SZ and optical positions (arcmin).\\
(This table is available in its entirety in a machine-readable form.)
}
\label{taba3}
\end{minipage}
\end{table*}

\section{12 Planck clusters not matched within a redshift difference of $0.05(1+z)$}
\label{plancknot}

We find that 10 of 12 un-matched {\it Planck} clusters have a large
uncertainty of redshift and may be matched within a lager redshift
difference from our clusters. We present some details if anyone
concerns:

PSZ2 G181.71$-$68.65 has a redshift 0.1529 in the {\it Planck}
catalogue. Along the line of sight, we find a cluster at $z=0.3176$,
which has counterpart cluster with a photometric redshift of
$z=0.3206$ in the CFSFDP catalogue.

PSZ2 G215.19$-$49.65 has a redshift of 0.2399 in the {\it Planck}. The
redshift is 0.3678 in our catalogue and 0.3607 in the CFSFDP catalogue.

PSZ2 G224.53$-$30.27 has a redshift of 0.2001 in the {\it
  Planck}. Along the line of sight, we identify two clusters,
J051657.9$-$223705 and J051705.8$-$223654. The cluster,
J051657.9$-$223705, has a redshift of 0.3097 and a richness of 81.26,
which has a counterpart cluster at $z=0.2980$ in the CFSFDP
catalogue. The cluster, J051705.8$-$223654, has a redshift of 0.9844
with a lower richness of 30.46. Thus, PSZ2 G224.53$-$30.27 is likely
matched by the richer cluster at about $z=0.30$.

PSZ2 G307.72$-$77.87 has a redshift of 0.4530 in the {\it Planck}. The
redshift is 0.6030 in our catalogue and is 0.5995 in the CFSFDP
catalogue.

PSZ2 G261.09$-$66.14 has a redshift of 0.56 in the {\it Planck}. The
redshift is 0.6418 in our catalogue and is 0.6079 in the CFSFDP
catalogue.

PSZ2 G250.59$-$25.03 has a redshift of 0.55 in the {\it Planck}. The
redshift is 0.6332 in our catalogue and is 0.6249 in the redMaPPer
catalogue.

PSZ2 G262.01$-$32.98 has a redshift of 0.23 in the {\it Planck}. Along
the line of sight, we identify two clusters, J053327.2-542017 and
J053304.2-542345.  The cluster, J053327.2-542017, has a redshift of
0.2916 and a richness of 33.16, which has a counterpart cluster at
$z=0.3040$ in the redMaPPer catalogue. The cluster, J053304.2-542345,
has a redshift of 1.0330 and a larger richness of 33.19. Thus, PSZ2
PSZ2 G262.01$-$32.98 is contributed from both optical clusters.

PSZ2 G260.76$-$31.27 has a redshift of 0.25 in the {\it Planck}. The
redshift is 0.3172 in our catalogue and is 0.3141 in the redMaPPer
catalogue.

PSZ2 G268.51$-$28.14 has a redshift of 0.46 in the
{\it Planck}. The redshift is 0.3643 in our catalogue.

PSZ2 G281.09$-$42.51 has a redshift of 0.38 in the {\it Planck}. The
redshift is 0.5772 in our catalogue and is 0.5550 in the RASS
catalogue.

Along the line of sight of two clusters, PSZ2 G212.25$-$53.20 and PSZ2
G255.64-25.30, no clusters are found in our catalogue with a detection
SNR and a richness above the thresholds.

\end{appendix}

\label{lastpage}
\end{document}